# Microstructure and Water Absorption of Ancient Concrete from Pompeii: An Integrated Synchrotron Microtomography and Neutron Radiography Characterization

Ke Xu [a,b,*], Anton S. Tremsin [c], Jiaqi Li [a,*], Daniela M. Ushizima [b,d], Catherine A. Davy [e], Amine Bouterf [f], Ying Tsun Su [a], Milena Marroccoli [g], Anna Maria Mauro [h], Massimo Osanna [i], Antonio Telesca [g], Paulo J. M. Monteiro [a]

[a] Department of Civil and Environmental Engineering, University of California, Berkeley, CA, 94720, USA
[b] Computational Research Division, Lawrence Berkeley National Laboratory, Berkeley, CA, 94720, USA
[c] Space Sciences Laboratory, University of California, Berkeley, CA, 94720, USA
[d] Berkeley Institute of Data Science, University of California, Berkeley, CA, 94720, USA
[e] Univ. Lille, CNRS, Centrale Lille, ENSCL, Univ. Artois, UMR 8181 - UCCS - Unité de Catalyse et de Chimie du Solide, Lille F-59000, France
[f] Laboratoire de Mécanique et Technologie (LMT), ENS Paris-Saclay, CNRS Université Paris-Saclay, Cachan Cedex, 94235, France
[g] School of Engineering, University of Basilicata, Potenza 85100, Italy
[h] Head of research and innovation area of Archeological Park of Pompeii, via Plinio 4, Pompeii, NA, 80045, Italy
[i] General Director of Archeological Park of Pompeii, via Plinio 4, Pompeii, NA, 80045, Italy

* Corresponding authors.

E-mail addresses:

Ke Xu: ke_xu@berkeley.edu

Jiaqi Li: jiaqi.li@berkeley.edu

**Abstract**

There is renewed interest in using advanced techniques to characterize ancient Roman concrete due to its exceptional durability and low-carbon footprint. In the present work, samples were drilled from the "Hospitium" in Pompeii and were analyzed by synchrotron microtomography (μCT) and neutron radiography to study how the microstructure, including the presence of induced cracks, affects their water adsorption. The water distribution and absorptivity were quantified by neutron radiography. The 3D crack propagation, pore size distribution and orientation, tortuosity, and connectivity were analyzed from μCT results using advanced imaging methods. Porosity was also measured by mercury intrusion porosimetry (MIP) as a reference. Ductile fracture patterns were observed once cracks were introduced. Compared to Portland cement mortar/concrete, the Pompeii samples had relatively high porosity, low connectivity, and a similar coefficient of capillary penetration. In addition, permeability was predicted from models based on percolation theory and pore structure data to evaluate the fluid transport properties. Understanding the microstructure of ancient Pompeii concrete is important because it could inspire the development of modern concrete with high durability.

**1. Introduction**

The desire to extend the life cycle of modern concrete infrastructure and to reduce its lifespan carbon footprint has instigated renewed interest in determining the reasons for the long-term durability of ancient Roman concrete. The production of ancient Roman concrete, which was made of volcanic ash, lime, water and volcanic rocks, has a lower carbon footprint than the production of Portland cement concrete, which is responsible for ~5% of global anthropogenic $CO_2$ emissions [1]. Moreover, the fact that some of the Roman marine concrete structures, including seawalls and harbor piers, have remained in good condition for 2,000 years, either fully immersed in seawater or partially immersed in shoreline environments, attests to the remarkable durability of ancient Roman concrete [2, 3]. Thus, the composition and microstructure of ancient Roman concrete can inspire the development of modern concrete with a low carbon footprint and high durability.

State-of-the-art techniques that were developed to study high-performance materials have recently been used to characterize the composition and microstructure of various ancient Roman concrete samples. Synchrotron radiation methods have been especially successful in providing new insights into the complex micro- and nano-structure of ancient Roman concrete. The use of high-pressure X-ray diffraction has provided information about the mechanical behavior of Al-tobermorite, which forms in ancient Roman seawater concrete, while soft X-ray nanotomography has allowed for mathematical morphology quantification of the clusters of Al-tobermorite nanocrystals [4, 5]. Jackson et al. [3] produced seminal work on the formation of phillipsite and Al-tobermorite using X-ray microdiffraction. Scanning transmission X-ray microscopy (STXM) has proven to be a powerful tool for understanding the nanoscale bonding environments of Al and Si in Al-tobermorite and C-A-S-H [6]. Using synchrotron high-resolution X-ray diffraction (XRD), Vanorio and Kanitpanyacharoen [7] formulated a ground-breaking hypothesis that the caprock of the Campi Flegrei Caldera has a fibrous microstructure that is similar to the existing ancient Roman concrete formed by lime-pozzolanic reactions. Recently, Palomo et al. [8] published a critical evaluation of the current developments in the characterization of ancient Roman concrete and how it can contribute to the new generation of green hybrid cement. Moreover, there is a strong motivation to study the effect of time on modern hybrid cement/concrete based on the understanding of ancient Roman concrete.

There is a consensus that the lack of concrete durability is associated with the penetration of water, and concrete often contains deleterious species. This deterioration is exacerbated by the presence of cracks, which provide a faster pathway for the ingress of water. Jackson et al. [9] studied crack resistance and resilience in Imperial Roman architectural mortar, but further information on the effect of the microstructure is still needed, especially the effects of cracks on concrete water ingress. Therefore, one of the purposes of the present work is to provide quantitative data on the dynamics of water absorption by uncracked and cracked samples from Pompeii using neutron radiography and to provide 3D microstructural data of uncracked and cracked samples with synchrotron microtomography. The integration of X-ray microtomography and neutron imaging, implemented in this study, allowed a systematic investigation of how concrete microstructure affects the dynamics of water penetration in Pompeii concrete samples. The unique capability of neutron imaging to map water distributions within concrete samples is enabled by the relatively high neutron attenuation cross-section by hydrogen [10, 11], which has relatively low absorption by the concrete itself. Synchrotron X-ray microtomography (μCT) can determine the complex topology of the phases that exist in ancient Roman concrete. The extracted samples from Pompeii did not show any cracks, and thus, after the μCT scans, cracks were introduced by loading, and the samples were retested to measure the influence of cracks on the fluid flow. Due to the complexity of the microstructure, it was necessary to improve the existing segmentation methods for 3D tomographic images. While a separate publication describes the intricacies of the proposed methodology [12], a short description appears in section 2 for the benefit of the interested reader.

## 2. Samples and experimental methods

Currently, approximately 49 of the 66 hectares of Pompeii have been excavated. In 1858, the Archeologist Giuseppe Fiorelli [13] divided the city into "regions" (neighborhoods) and "insulae" (blocks) (see Fig. 1). The present work studies the Hospitium building (Fig. 2a) located at Stabiana street n.3 (Fig. 2b) near the Stabiae gate in insula 1 of Regio I (Fig. 2b). On the eastern side of Stabiana street, with the exception of very few accommodations, there was an almost uninterrupted series of tabernae hospitia, cauponae, stabula and officinae. The Hospitium building, excavated in 1872, has an area of approximately 600 square meters and is delimited to the North by a shop (street n.4), to the South by an old "popina" (Roman wine bar/tavern frequented by the lower classes and slaves, where a limited menu of simple foods and wines was available, street n. 2), to the East by "Via Stabiana" and to the West by a nonpaved alley. The ground floor had a large stable and spacious "latrina"; the first floor, most likely accessible through a staircase, was covered and composed of only bedrooms.

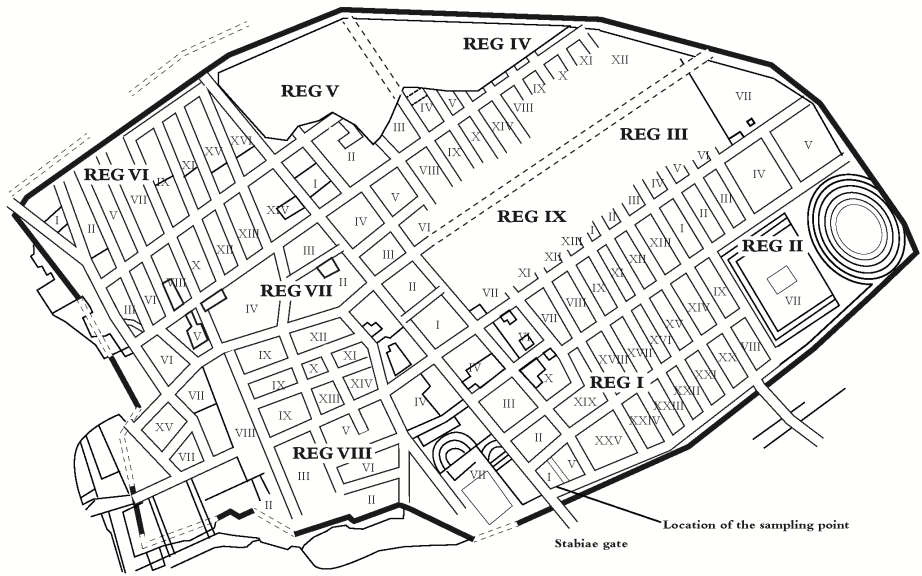

**Figure 1.** Plan of Pompeii with the location of the sample extraction.

All mortar samples were carefully collected at a depth of approximately 50 mm on both lateral sides of the building walls and pillars to avoid weathering of any type. A small core drill was employed for extracting larger samples from the walls of the house, and both a hammer and a small chisel were used for collecting the bedding mortars between bricks.

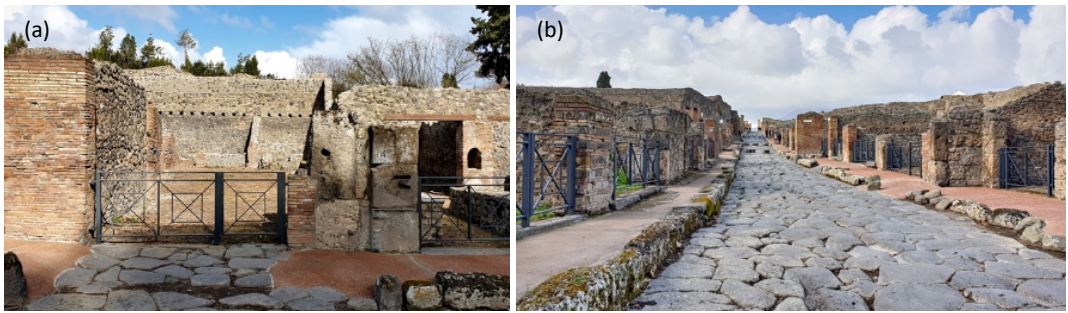

**Figure 2.** Photographs of (a) Hospitium entrance and (b) Via Stabiana.

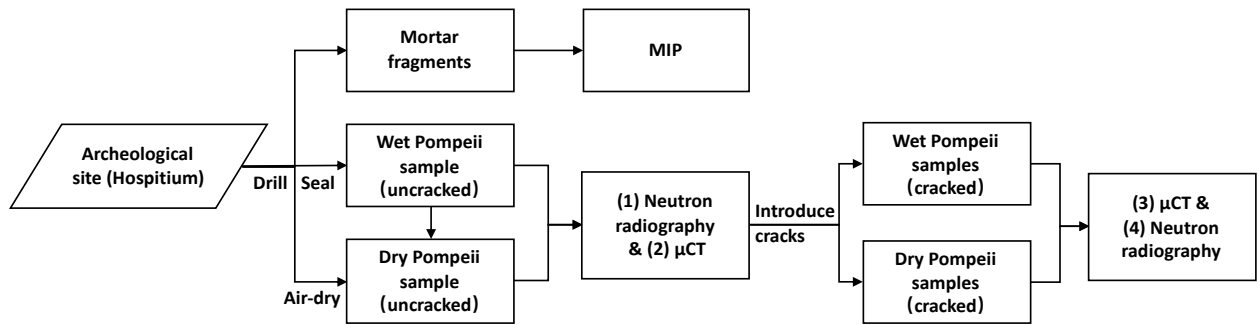

**Figure 3**. Schematic diagram of the experimental workflow for the study of ancient concrete from Pompeii.

A series of experiments (see Fig. 3) was conducted to systematically investigate the microstructure and water transport mechanism in the mortar samples. The numbers indicate the sequence of the experiments, and detailed information about each experiment is described in the following sections.

Three mortar fragments were investigated with mercury intrusion porosimetry (MIP). The porosity measurements were performed using a Thermo-Finnigan Pascal 240 Series Mercury Porosimeter (maximum pressure, 200 MPa) equipped with a low-pressure unit (140 Series) that could generate a high vacuum level (10 Pa) and operate between 100 and 400 kPa. With increasing pressure, mercury is gradually able to penetrate the bulk sample volume. If the pore system is composed of an interconnected network of capillary pores in communication with the outside of the sample, mercury enters the smallest pore neck at the corresponding pressure value. If the pore system is discontinuous, mercury could penetrate the sample volume provided that its pressure is sufficient to break through the pore walls. In any case, the pore width related to the highest rate of mercury intrusion per change in pressure is known as the ''critical'' or ''threshold'' pore width [11], which represents the lowest width of pore necks that connect a continuous system. A unimodal or multimodal pore size distribution can be obtained, depending on the occurrence of one or more peaks in the derivative volume plot.

2.1 Neutron radiography and water (capillary) absorption experiments

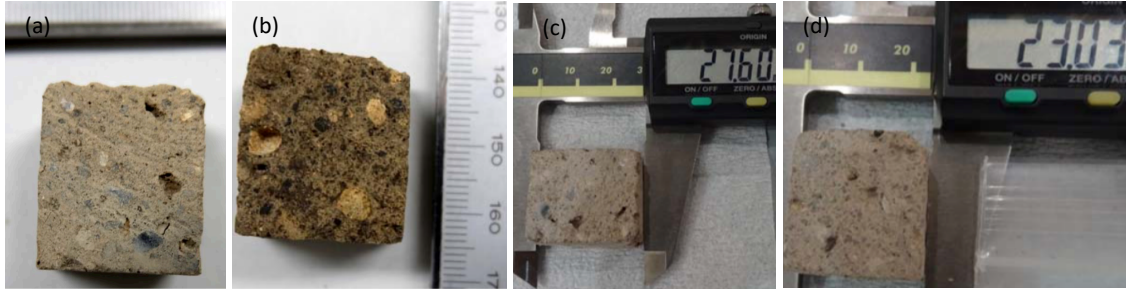

**Figure 4.** Geometry and dimensions of two samples: (a) uncracked DP, (b) uncracked WP, (c) cracked DP, (d) cracked WP.

Two concrete samples from the same location in Hospitium, Pompeii, with dimensions of 23.5x27.6x27.6 mm$^3$ and 17.8x23.0x27.5 mm$^3$, were studied. The dry Pompeii sample (DP) was air-dried after drilling from Hospitium, which is located along Via Stabiana, Pompeii, while the wet Pompeii sample (WP) was originally wet and sealed after drilling to analyze the original water distribution in ancient Roman concrete. For air-drying conditions, the sample was kept under a vacuum in a desiccator that contained silica gel and soda lime to ensure protection against $H_2O$ and $CO_2$. First, both samples (Fig. 4 a and b) were scanned using neutron radiography and synchrotron microtomography to analyze the water distribution, water absorption, and microstructure at the original state (uncracked state). After these measurements, cracks were introduced into both samples by applying uniaxial compressive stresses. Finally, both cracked samples (Fig. 4 c and d) were scanned again using neutron radiography and synchrotron microtomography to compare the water distribution, water absorption, and microstructure measured at the original state with the measurements obtained at the cracked state.

Neutron imaging experiments were performed at the Materials and Life Sciences (MLF) Facility at the Japan Proton Accelerator Research Complex J-PARC. The pulsed neutron beam at this spallation neutron source operated at 25 Hz neutron pulse frequency and allowed the measurement of neutron transmission spectra in every pixel of the images. More than 250,000 transmission spectra were measured in the experiments simultaneously, each within 55 μm pixels of the 512x512 neutron counting detector. A fast neutron counting detector with microchannel plates (MCPs) and Timepix readout with an active area of 28x28 mm$^2$, built at the University of California at Berkeley in collaboration with Nova Scientific, was used for the energy-resolved neutron radiography measurements [14-16].

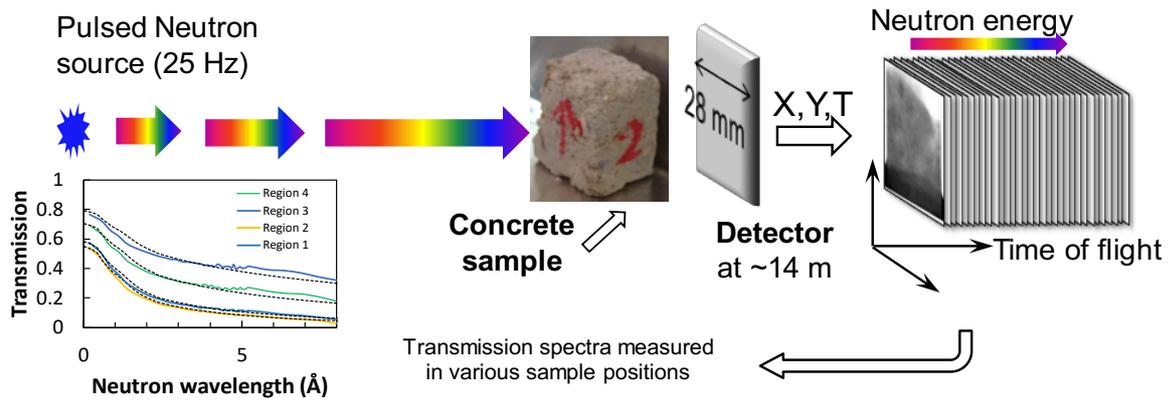

**Figure 5.** Schematic experimental setup of the neutron energy-resolved imaging used for mapping the water distribution and dynamics of water penetration.

The neutron time of flight (ToF) method was used to reconstruct the energy of each detected neutron, allowing transmission spectra to be measured simultaneously in a wide range of energies from epithermal (~eV energy) neutrons to cold neutrons (meV energies). All of the measured spectra were normalized by the spectrum taken with no sample present in the beam to correct for the spectral features of the interrogating beam and the spatial nonuniformity of both the detector response and the neutron beam itself.

The analysis of these neutron transmission spectra allowed quantitative mapping of the water distribution in as-extracted concrete samples as well as investigation of the dynamics of water penetration in as-extracted and cracked samples. More details on the reconstruction of water concentration in concrete samples are described in [15, 17]. The samples were placed 5-6 cm from the active area of the detector, which resulted in only a small fraction (<2.5%) of neutrons to be scattered by water to reach the detector, and it introduced unwanted background signals. Because of the small contribution of scattered neutrons, no scatter rejection techniques were implemented in the present study. The accuracy of the water concentration mapping was verified with a calibration step wedge sample where the known thickness of water was reconstructed from the measured neutron transmission data [17].

The dynamics of water penetration were determined with a white spectrum beam to map the location of water in the sample on a subsecond time scale. Measurements of neutron transmission spectra in each pixel of the data set require relatively long integration (tens of minutes) to acquire sufficient statistics, while the dynamics of water absorption were studied with multiple images acquired per second. Water was introduced into the bottom part of an aluminum container, in which the concrete samples were placed in front of the detector, and consecutive transmission images were acquired multiple times per second. Thus, the location of the waterfront driven up into the sample by capillary forces was visualized as a sharp drop in the sample transmission at the boundary of wetted and dry concrete samples.

2.2 Synchrotron X-ray microtomography experiments

Synchrotron μCT was performed at beamline 8.3.2 at the Advanced Light Source (ALS) of the Lawrence Berkeley National Laboratory [18]. A white beam was used in the experiment, where the X-ray beam needs to go through the concrete samples. The beam energy was set to 35 keV with a constant beam current of 500 mA. During a scan, the sample was mounted on the circular holder and was rotated about an axis perpendicular to the horizontal plane over 180°, and 1969 2D radiographs in total were obtained. Each projection was acquired on a 2560 px CCD camera (PCO edge) equipped with an 8.3 mm field of view and a 1x Mitutoyo magnification optical objective lens with a spatial resolution of 9.5 μm. To improve the phase contrast and achieve better image quality, an exposure time of 15 ms per projection was used to keep the transmission ratio of the beam below 40% based on the Beamline 8.3.2 Manual. After each scan, a 3D data volume (tomogram) was reconstructed from the 2D radiographs using the filtered back projection algorithm on TomoPy [19, 20]. For each sample, two stacks of tomograms (uncracked and cracked states) with dimensions of 2560*2560*2324 voxels$^3$ (32-bit 56.7 GB for each stack) were obtained for analysis.

2.3 Preprocessing, visualization and segmentation of tomography images

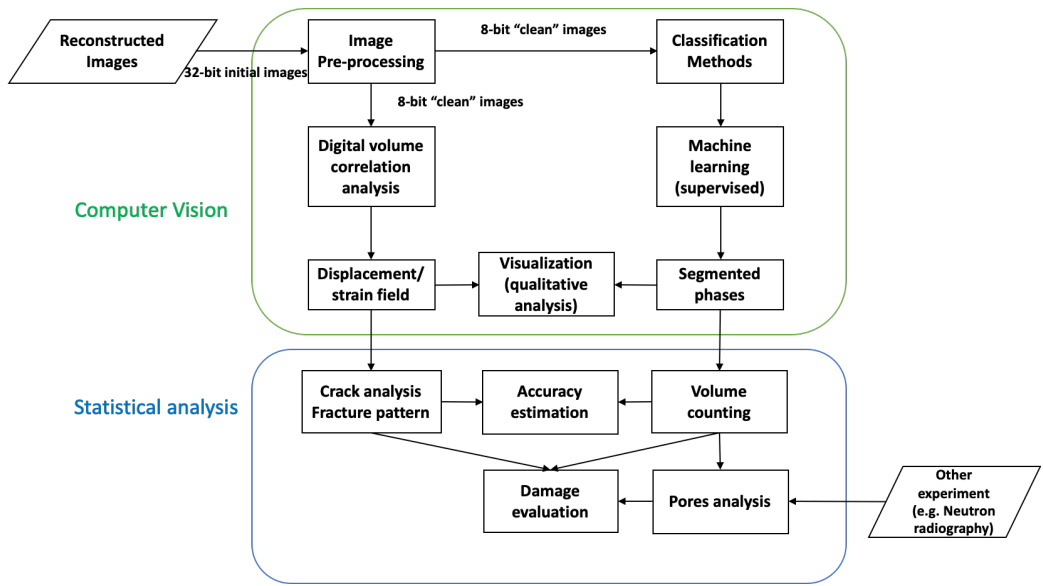

**Figure 6.** Schematic diagram showing the stages of computer vision and statistical analysis for tomography images

To conduct a systematic investigation of the sample microstructure and combine neutron radiography images, we built a new pipeline (Fig. 6) for autonomous and accurate μCT image analysis. This pipeline contained two parts: computer vision and statistical analysis. In the computer vision analysis, preprocessed μCT images were segmented into different phases using machine learning algorithms. The edge-preserving filters, mathematical morphology and an assortment of supervised machine learning algorithms, such as random forest[12], improved the segmentation accuracy when compared to the traditional grayscale threshold segmentation method. At the same time, advanced digital volume correlation (DVC) algorithms were performed on the preprocessed μCT images of reference samples and deformed samples to calculate the displacement map and strain field. Both segmented 3D pore phase images and 3D principal strain fields revealed cracking propagation. The DVC results also allowed us to qualitatively verify the accuracy of the phase segmentation. Then, qualitative and quantitative analysis was conducted based on 3D segmented phase images and strain fields. In addition, a comprehensive microstructure investigation was conducted that combined neutron radiography and segmented images.

Image preprocessing was performed on Fiji [21], and this stage involves the following: a) transforming the slice images from 32-bit to 8-bit, b) removing the overlap (20 was set during the tile mode scans) of images between image stacks, c) removing the partial background and cropping out the region of interest (ROI), d) eliminating noise and ring artifacts from the images through 3D bilateral filtering [22], and e) improving the contrast with a saturated pixel of 0.35. For DP and WP, two stacks of images could have different 3D rendering orientations and positions since the uncracked and cracked samples were scanned separately. The image volume registration method was applied to transform two stacks of images into one coordinate system for one sample. Translation transformation was used to preserve the same coordinates for the rotation centers of two stacks. Then, the cracked stack was rotated to keep the orientation of the bottom image slice consistent with that of the uncracked stack, which reduced the rotation artifacts between scans. In this study, Tomviz [23] was used to visualize the reconstructed 3D μCT images, segmented 3D μCT images of different phases, and 3D crack evolutions from DVC. Dragonfly (Object Research Systems (ORS) Inc.) was used to visualize the 3D connectivity of pore networks.

Image segmentation is the most crucial step for the quantitative analysis of 3D μCT. To obtain the morphological representation of specific phase systems (e.g., the porous system in concrete) and conduct further statistical analysis (e.g., pore size distribution), a segmentation process is commonly implemented to separate images into discrete phases (e.g., pore phase, aggregate phase, and matrix phase in concrete). Local and global grayscale thresholding are the most commonly applied approaches for CT of concrete [24, 25]. The major drawback of these methods is to utilize one feature (i.e., the grayscale in the images), and their accuracy and robustness are limited by the quality of the tomographic images [26, 27]. The segmentation results are unstable for tomographic images, with low phase contrast, noise points, background regions, or attenuation problems.

Recently, machine learning (ML) algorithms have been successfully applied to the segmentation of structures with X-ray μCT images of mineralogical samples. Shipman et al. [28] extracted quantitative mineralogical information with regard to the composition, porosity, and particle size of chromite ore samples. Camalan et al. [29] estimated the 2D mineral map and its associated liberation spectrum of a particular chromite sample by using random forest classification. These ML classifiers are robust, automatic, and multifeature based. For the combination of extracting multiple image features from the raw data and applying ML image segmentation algorithms, the Trainable Weka Segmentation Tool [21, 30] was used in this study. Considering both the image segmentation accuracy and computing time, grayscale, mean, median, minimum, maximum, Gaussian blur, Hessian, anisotropic diffusion, and difference of Gaussian were selected as the features for segmentation. At the same time, a small number of images was randomly selected from the whole data set as the training images. After the feature selection, different regions (the most obvious objects that corresponded to the target phases) in the training images were annotated as target phases, which constructed the training data set.

Five phases were separated from each other in the μCT of concrete: background, supporting stage for the test, aggregate, matrix, and pore. Cracks were included in the pore phases. Different classifiers were trained using the selected features and training data set as a preliminary step. Based on the initial segmentation results on the training data set, more annotations were added to the misclassified regions, and the classifiers were retrained and tuned until satisfactory segmentation results (testing pixel accuracy above 90% for each phase) were obtained. After evaluating the trained models, the most suitable ML classifier was selected based on the segmentation results. The random forest classifier, proposed by Tin Kam Ho [31], proved to have better performance when compared with other ML classifiers (e.g., linear perceptron classifier, SVM classifier, and Gaussian discriminant classifier) on our data set. In addition, the random forest algorithm has been one of the most accurate learning algorithms for μCT segmentation, and it produces a highly accurate classifier for many CT image data sets [32]. Therefore, the random forest classifier was selected and applied to the μCT dataset of DP and WP. Fig. 7 shows the segmentation procedures from a μCT image to the segmented mask of the pore phase using the random forest classifier. Table 1 shows the confusion matrices that highlight the accuracy in pixel classification results of the applied random forest model with inputs obtained from manually segmented samples (manual annotations). Note that both supporting phase and background phase were regarded as background phase in the construction of confusion matrices. Ushizima et al. [12] described several of these algorithms in detail using Python scripts to characterize the microstructure based on μCT image segmentation from similar ancient Roman concrete samples.

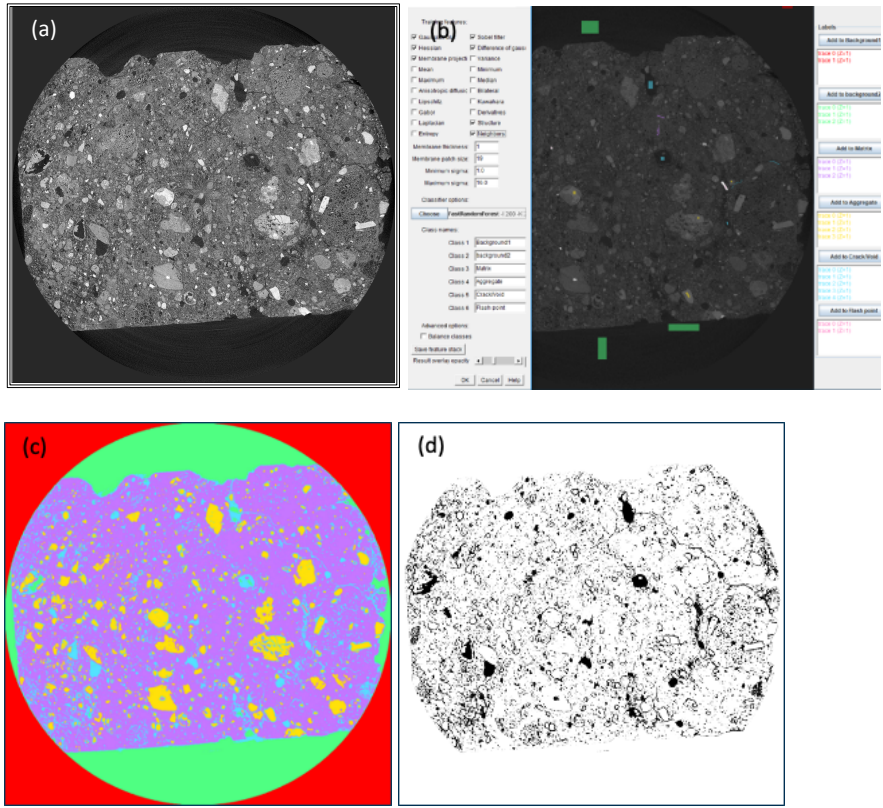

**Figure 7.** (a) Initial image, (b) feature selection and image annotations, (c) final classified image (Red phase: background, Green phase: supporting stage, Yellow phase: aggregate, Purple phase: matrix, Blue phase: pores), (d) segmentation of pores.

**Table 1**. Confusion matrix of the testing dataset with random forest classification.

| %          | Background | Aggregate | Matrix    | Pore      |
|------------|------------|-----------|-----------|-----------|
| Background | **97.52**  | 0         | 0         | 2.48      |
| Aggregate  | 0          | **96.82** | 3.18      | 0         |
| Matrix     | 0.01       | 0.17      | **98.43** | 1.39      |
| Pore       | 0          | 0         | 5.21      | **94.79** |

2.4 Digital volume correlation

Separating intrinsic voids and introduced cracks from 3D μCT with high accuracy depends upon image segmentation algorithms [33-35] and data resolution. DVC is an image processing method that is often applied in the analysis of crack propagation without the distraction of intrinsic voids for the microtomography test. Roux et al. [36] calculated the displacement fields and evaluated the accuracy of the algorithm based on the analysis of an in-situ uniaxial compression CT test on a solid polypropylene foam. Bouterf et al. [37] analyzed the crack propagation and degradation mechanisms of lightweight plasterboard via a nail pull test conducted in-situ in laboratory microtomography. DVC techniques calculate the displacement field by minimizing the correlation residuals, i.e., the sum of squared grayscale differences between the reference volume and the deformed volume corrected by the calculated displacement field [37], under the hypothesis of conservation of the grayscale of the analyzed microstructure of images. Then, the strain fields can be calculated from the displacement fields, and the crack propagation can be visualized and quantified [38, 39].

In this study, the displacement fields were measured when discretized over a finite element mesh made of 8-node cubes (C8) [36]. Since the strains were relatively small, a mechanical regularized approach was used [40, 41]. A steep displacement gradient indicated the presence of a crack, and the maximum equivalent strain was a measure of the crack opening displacement magnitude, which revealed that a complex pattern of cracking was more sensitive than the segmented pore images. The C8R developed in MATLAB by Bouterf [42, 43], which used the regularized C8 element, was selected to conduct the DVC analysis for both DP and WP. Due to the limited storage of MATLAB, the original reconstructed µCT stacks ($2560*2560*2324$ voxels$^3$) were rescaled to $1280*1280*1162$ voxels$^3$. For the mesh generation, the element size, which is comparable to the zone of interest (ZOI) size of classical digital image correlation codes, was optimized to be $20*20*20$ voxels$^3$ for global DVC to reduce the computational complexity and uncertainty while still being able to capture complex displacements and microcracks [40, 41]. The regularized lengths were set to be equal or larger than the element size in such a way that the high-frequency displacement fluctuations, which are not mechanically admissible, are filtered out.

2.5 Pore morphology and statistical analysis

3D pore network information is essential to assess the physicochemical properties of concrete, such as mechanical strength and permeability. 3D pore morphology and statistical analysis were conducted on segmented pore phase images (binary images). More accurately, the pore size distribution, porosity, pore area distribution along the Z-axis, aspect ratio, pore connectivity, connectivity density, and tortuosity were characterized based on 3D segmented images of the pore phase.

The 3D discrete pore size distribution (DPSD) and continuous pore size distribution (CPSD) were computed similarly to the work by Münch et al. for pore analysis [44]. DPSD is a simple measurement, where each pore object is considered to be a sphere with the same volume as the original pore. DPSP is the relative pore volume as a function of the equivalent sphere diameter [44-46]. In comparison, the CPSD is based on the assumption that the pore network is a continuum and can be invaded by fluid as in MIP [44]. In the CPSD measurement, the pore objects (defined as spheres of identical volumes) are invaded from the largest pores to the smallest pores without the "ink bottle effect". CPSD is defined as the relative volume of invaded spheres as a function of the sphere size [44]. In this paper, 3D DPSD and CPSD were performed by using the Xlib plugin [44] in Fiji.

Porosity is defined as the percentage of pore voxels in the volume of interest (VOI). The cross-sectional areas of the pores and cracks along the height (Z-axis) were calculated from 3D binary images of the pore phase to associate the water absorption results from neutron radiography. Voxel connectivity analysis was performed based on a 6-connected voxel criterion, which means voxels are considered connected to form the "cluster/object" when the voxels share a common face with each other (pores are usually made of at least two voxels). Then, the pore objects (clusters) were used to characterize the aspect ratio, pore connectivity, and connectivity density.

To describe the morphology of pore objects in 3D, the aspect ratio (elongation) of each individual 3D pore object is defined as the ratio between the smallest eigenvalue and the largest eigenvalue for the inertia eigenvectors, where inertia eigenvalues and eigenvectors are calculated from the inertia tensor (equivalent ellipsoid) of the 3D pore object. Therefore, the aspect ratio ranges between 0 and 1, where a value of 1 corresponds to a sphere. Both pore connectivity and connectivity density indicate the degree of connectivity in the pore network, but their definitions are different. Pore connectivity (expressed in %) is defined as the number of void voxels in the largest percolating pore object (cluster) divided by the total number of voxels attributed to pores in the VOI [47, 48]; it is a fraction of the porosity and is equal to 100% when all of the pores in the system are percolating. For uncracked samples, no percolating pore system is obtained for the whole image stack. In such an instance, the pore connectivity is the measurement of the percolating pore object limited to the lower half-height of the sample.

In comparison, the measurement of connectivity density required further computations with the BoneJ plugin [49-51] in Fiji. The pore object (network) was first skeletonized with a topology-preserving medial axis algorithm that allowed branch (i.e., trabecule) and junction analysis [52]. Then, the number of connected trabecules in the VOI was measured, and the connectivity density (expressed in um$^{-3}$) was defined as the number of trabecules divided by the VOI [50, 51]. The tortuosity was quantified by generating random walkers within the skeletonized pore network and recording their traveled free distance with the AnalyzeSkeleton [53] plugins in Fiji. The tortuosity is defined as the mean free distance divided by the Euclidean distance from the start point to the end point [53-55]. In this paper, VOI refers to the whole sample volume since a global analysis was conducted.

## 3. Results and discussion

3.1 Porosity Measured by Mercury Intrusion Porosimetry

The reproducibility of mercury porosimetry measurements among the three replicates was successful; in fact, the greatest total porosity value was 36.2%, while the lowest was 34.5%, which corresponds to a total cumulative intruded Hg volume equal to 248.4 and 217.7 mm$^3$/g, respectively. As an example, Fig. 8 reports the cumulative and derivative Hg volume for one of the three inspected samples; it can be noted that there was a very sharp unimodal pore size distribution centered on a threshold pore radius equal to 0.79 μm (a value related to the other two mortar fragments, which were 0.77 and 0.86 μm). According to the MIP results from Herculaneum-Roman mortars [56], the examined samples belong to the "medium degree of porosity (20-40%)" group. Notably, the measured pore sizes in these mortars are more than one order of magnitude larger than the usual pore sizes of the percolating pore system of modern Portland cements (C-A-S-H pores), and the latter are reputed to be in a 10-40 nm range [57-59]. Such small pores are present in non-negligible amounts in the pore size distribution (PSD) given in Fig. 11, although they do not correspond to the peak size. The peak pore sizes of 0.77-0.86 μm can be attributed to paste damage due to Hg pressure or to sample damage during its preparation for Hg porosimetry.

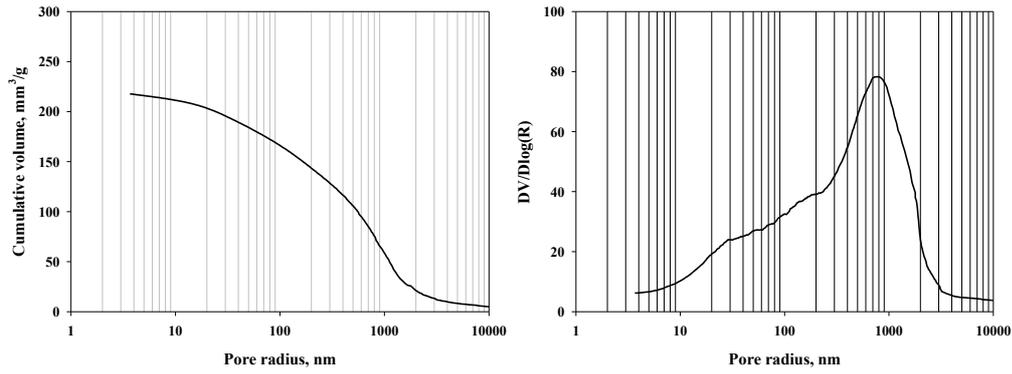
**Figure 8**. Cumulative (left) and derivative (right) Hg volume vs. pore radius for the mortar from the pillar.

3.2 Microtomography results

3.2.1 Microstructure analysis of dry Pompeii sample before and after introducing cracks

The direct 3D renderings of reconstructed tomograms (Fig. 9 a and b) often generate views that occlude the complicated 3D cracking patterns in concrete, especially the internal cracks. The segmented pore phases (Fig. 9 e and f) were rendered to reveal the differences between the pore networks. In the region $\{X, Y, Z: 12.5$ mm$\leq X \leq 20$ mm, $0.1$ mm$\leq Y \leq 10$ mm, $3$ mm$\leq Z \leq 25$ mm $\}$, new pore phases (white voxels) emerged due to the propagation of multiple cracks and formed a triangular fracture plane after the mechanical compression test. By comparing the cracked pore phase and aggregate phase, most cracks propagated along the edges of aggregates. This result suggests that the interface transition zones (ITZs) in DP were the weak zones where crack initiation and propagation occurred. Further segmentation could be performed to visualize the introduced cracks without the initial pores, but it is very challenging to further separate them due to the similarities of their gray level, particle size, spatial shape, and spatial distribution.

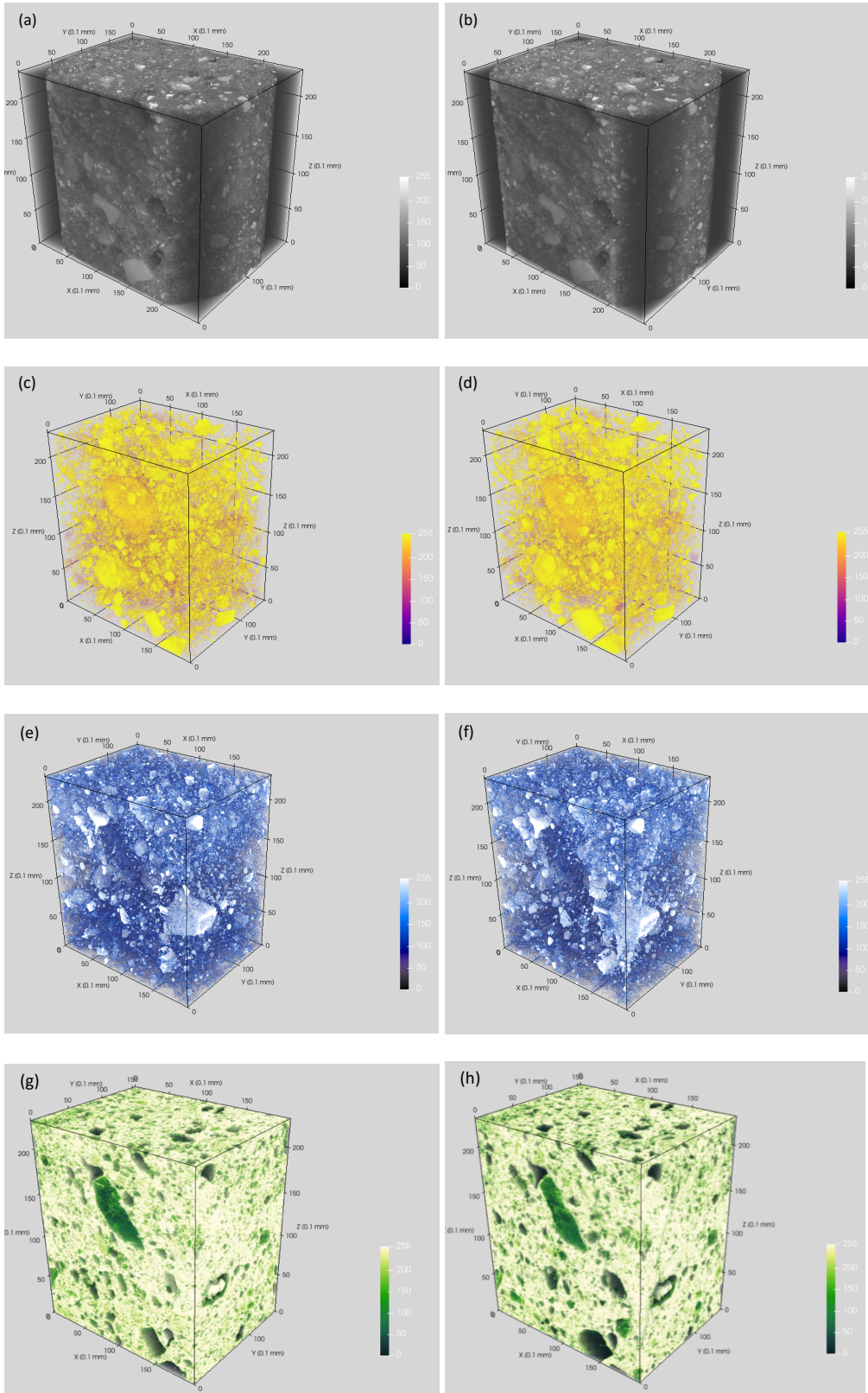

**Figure 9.** 3D rendering of reconstructed tomography and segmentation results for DP: (a) uncracked DP, (b) cracked DP, (c) aggregate phase of uncracked DP, (d) aggregate phase of cracked DP, (e) pore phase of uncracked DP, (f) pore phase of cracked DP, (g) matrix phase of uncracked DP, (h) matrix phase of cracked DP. Note that all of the scale bars represent the grayscale, and the segmented phase images are binary.

DVC analysis provides both a qualitative understanding of the complex 3D fracture patterns and a quantitative evaluation of the damage. The 3D displacement maps, major principal strain field, and residual error field (Fig. 10) were calculated using C8R DVC as described in the digital volume correlation section. For the color bar, the bright value indicates a high residual error, displacement, or strain. There are few residual errors, and the maximum error is under 0.1, which indicates a high accuracy of the DVC calculation. In the displacement maps, a steep strain gradient indicates the presence of a crack. In the principal strain field, the maximum principal strain is a measure of the crack opening displacement magnitude.

From the visualization of the 3D fracture pattern (Fig. 10), sample DP presents a ductile fracture pattern and consumes more energy than the brittle fracture pattern. In ordinary Portland cement concrete, crack paths preferentially develop in the porous ITZ zone between the fine-grained cement paste and the largely inert sand and gravel aggregate; crack paths usually have only a few microcracks and one macrocrack, and the sample presents a brittle fracture mechanism. However, the fracture behavior of DP, similar to the Type 5 failure of a typical cylindrical concrete during uniaxial compression (ASTM C39) [60], introduced extensive microcrack and macrocrack propagation. Type 5 fractures occurred due to the unbonded caps during the loading, and the ultimate capacity of the specimen might not have been attained. The multiple micro/macro crack propagations and the wider dispersion of cracks caused a load redistribution. Additional energy was absorbed by the diffuse networks of microcracks and macrocracks compared to the standard Type 5 fracture with a single localized macrocrack. By comparing Fig. 10 f and Fig. 9 f, the 3D fracture pattern from DVC analysis and the newly emerged pore phases provided similar results on the crack paths. The macrocrack propagations appeared in the top corner region $\{X, Y, Z: 12.5 \text{ mm} \leq X \leq 20 \text{ mm}, 0.1 \text{ mm} \leq Y \leq 10 \text{ mm}, 3 \text{ mm} \leq Z \leq 25 \text{ mm}\}$ of DP. The morphology of the macrocracking networks formed a triangular fracture plane. In addition, the calculated normal directions of the fracture plane from the two methods were the same. The above results verified the accuracy of both the image segmentation and the DVC analysis.

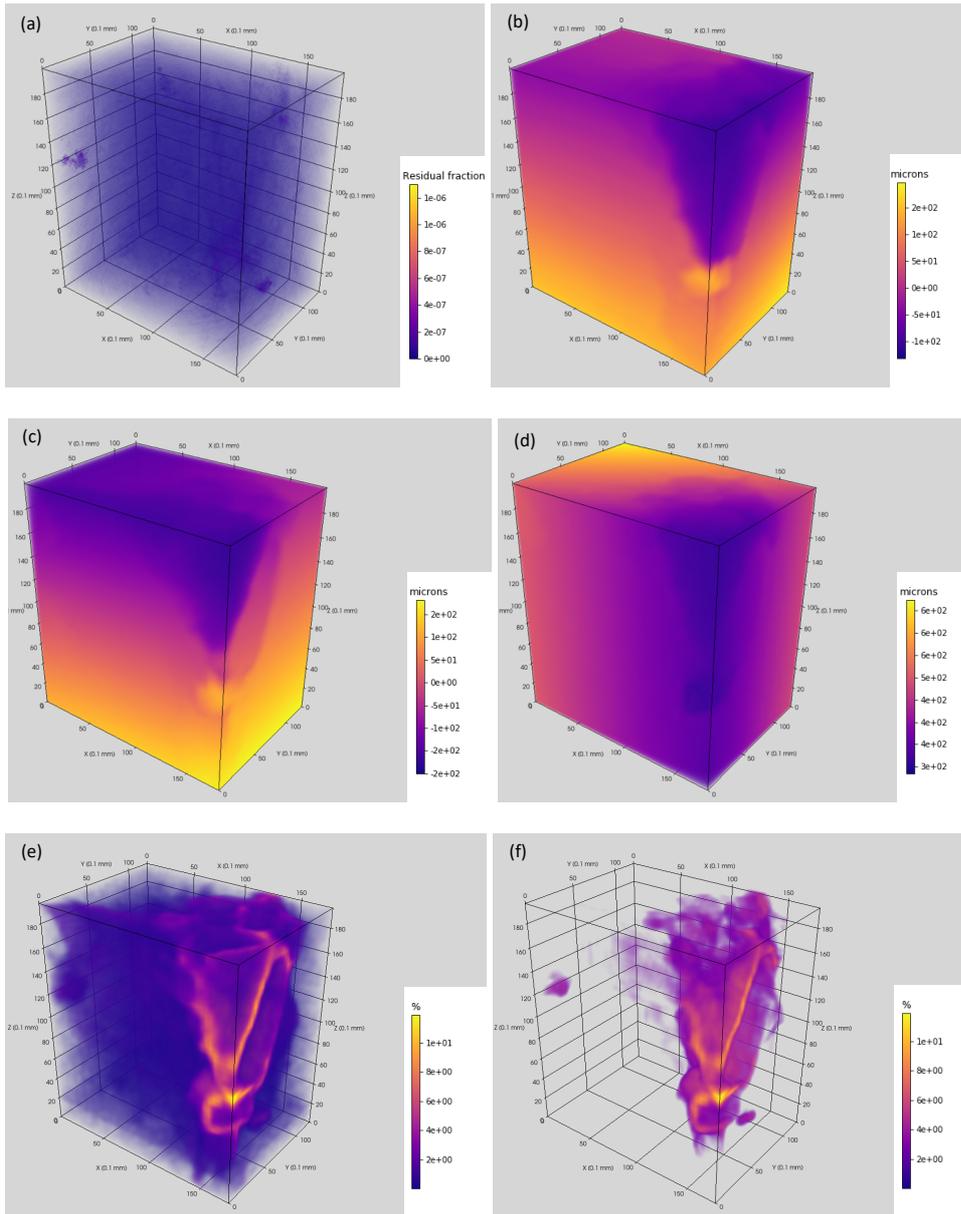

**Figure 10.** 3D renderings of (a) residual error field, (b) displacement Ux, (c) displacement Uy, (d) displacement Uz, (e) major principal strain $\varepsilon_{eq}$, and (f) visualization of cracking propagation in DP. The difference between (e) and (f) is that the strain fields below 1% were ignored in (f) to visualize the crack network and eliminate the noise.

3.2.2 Microstructure analysis of wet Pompeii sample before and after introducing cracks

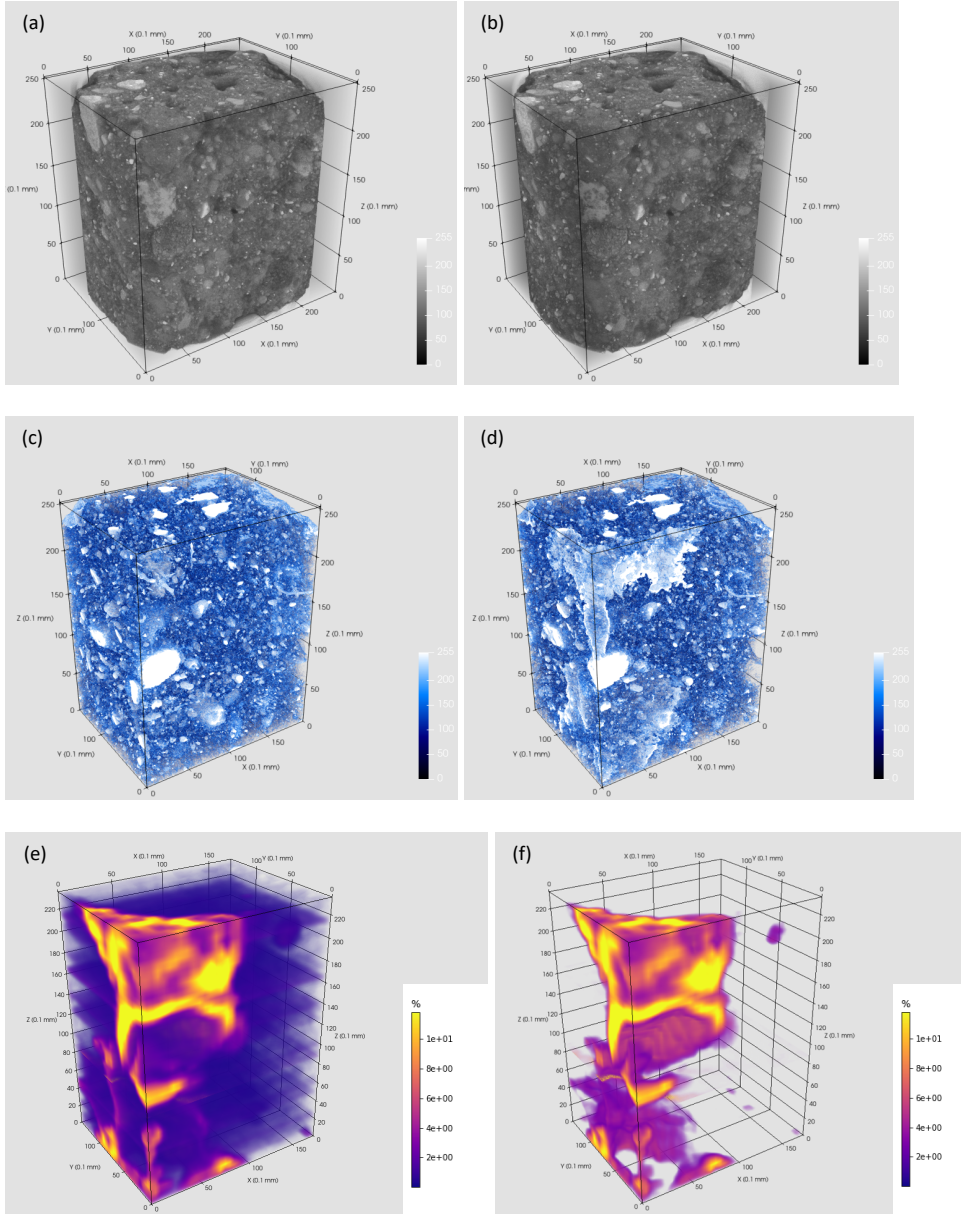

**Figure 11.** 3D rendering of reconstructed tomography, segmentation and DVC results for WP sample (a) Uncracked WP sample, (b) cracked WP sample, (c) pore phase of uncracked WP sample, (d) pore phase of cracked WP sample, (e) major principal strain $\varepsilon_{eq}$, (f) visualization of cracking propagation. The difference between (e) and (f) is that the strain fields below 1% were ignored in (f) to visualize the crack network and eliminate the noise.

Using the same segmentation, DVC analysis and visualization procedures were applied on the WP dataset (Fig. 11). The pore phase at both the uncracked and cracked states was rendered to reveal the microstructure of the WP sample. Similar to the DP sample dataset, the segmentation results of region $\{X, Y, Z: 2.5 \text{ mm} \leq X \leq 22.5$ mm, 0.1 mm ≤ $Y$ ≤ 19.9 mm, 0 mm ≤ $Z$ ≤ 25 mm} in the original coordinate system (Fig. 11 a and b) were rendered to avoid the misclassified background phase and stage phase. The WP sample also preserved the same aggregate size distribution after introducing the cracks. There is no obvious deformation or relative displacement of aggregate particles, which is the same as the DP sample. When comparing the pore phase in the uncracked WP sample and cracked WP sample, new pore spaces (white voxel) emerged in the region {X, Y, Z: 0 mm ≤ X ≤ 10 mm, 0 mm ≤ Y ≤ 10 mm, 0 mm ≤ Z ≤ 25 mm} where multiple cracks propagated. The cracking pattern of the WP sample is similar to that of the DP sample, but the WP has wider macrocrack propagation, and its cracks formed a large connected porosity network. In general, the wet Pompeii sample and dry Pompeii sample have similar fracture patterns. By comparing Fig. 11 f and Fig. 11 d, the 3D fracture pattern from DVC analysis and the newly emerged pore phases showed similar results for the crack paths. The macrocrack propagations occurred in the side corner region {X, Y, Z: 0 mm ≤ X ≤ 10 mm, 0 mm ≤ Y ≤ 10 mm, 0 mm ≤ Z ≤ 24 mm} of the WP sample, which corresponds to the porosity distribution in Fig. 11 d. The above results again verified the accuracy of the image segmentation. The multiple microcrack and macrocrack propagations and the broader dispersion of cracks in the WP sample are similar to the ductile cracking propagation and fracture pattern in the DP sample. Our previous paper extensively discussed how needle-like and fiber-like phases contribute to the mechanical resistance and durability [4, 6, 9, 61] of concrete.

3.2.3 3D statistical analysis

A more detailed cracking pattern analysis is conducted through 3D connectivity calculations for the pores and cracks. The segmented images of the pore phase were imported as ROI files into Dragonfly. Object analysis statistics were performed to calculate the volume, volume diameter, porosity, and pore (network) connectivity, as shown in Table 2. Note that the porosity of uncracked samples from μCT is much lower than the porosity obtained by MIP on the uncracked mortar fragments because μCT characterized only a fraction of the total porosity measured by MIP. According to the measurements from μCT alone, there is no percolating pore network in the whole volume of the uncracked samples obtained. This finding means that the fully percolating pore network in the uncracked Pompeii samples is below the resolution of μCT (below 9.5 x 2=19 μm), and hence, the fluid permeability is due to the pores that are smaller than 19 μm; using a typical 1D Katz-Thompson model for fluid permeability (with a porosity of 8.68 or 13.48%, a critical pore size of 19 μm, and a tortuosity of 1) [62], the model's result means that the permeability is smaller than $1-2 \times 10^{-13}$ $m^2$. The comparison of pore structures between ancient Pompeii concrete and modern Portland cement paste/concrete is discussed in section 3.4. After introducing cracks, both DP and WP samples have only a minor increase (< 0.5%) in porosity and a minor decrease in tortuosity but a significant increase (~10%) in pore connectivity. Because of the low pore connectivity without the largest pore cluster in cracked samples, this result suggests that the volume of the introduced cracks is small. Note that percolating pore networks are observed only after introducing cracks, which suggests that the detected percolating pore networks in the cracked samples are composed of both initial large pores and, most importantly, introduced cracks.

**Table 2.** Porosity and pore connectivity of samples DP and WP before and after cracking, as given by μCT.

| Concrete | Total porosity | Connectivity | Connectivity density (um^-3) | Tortuosity (Mean ± Standard deviation) |
|---|---|---|---|---|
| uncracked DP | 8.68% | 0% (9.37% at half height) | 2.30E-09 | 1.32 ± 0.43 |
| cracked DP | 9.04% | 21.22% | 3.73E-08 | 1.30 ± 0.41 |
| uncracked WP | 13.48% | 0% (26.74% at half height) | 8.50E-09 | 1.26 ± 0.37 |
| cracked WP | 13.68% | 39.40% | 9.05E-08 | 1.25 ± 0.38 |

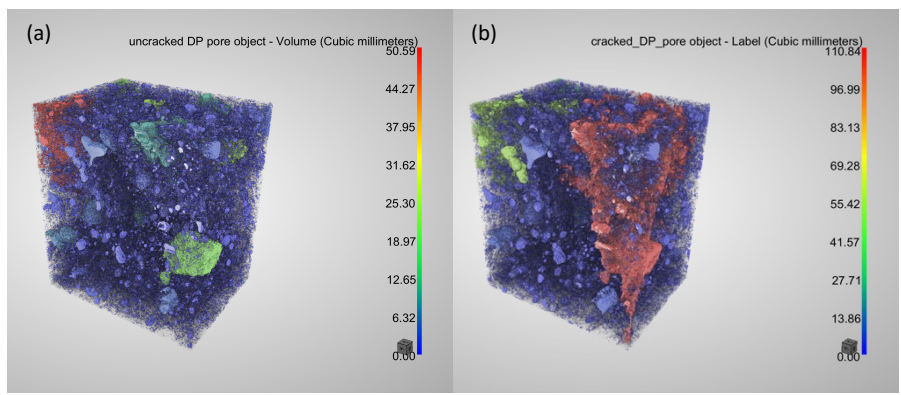

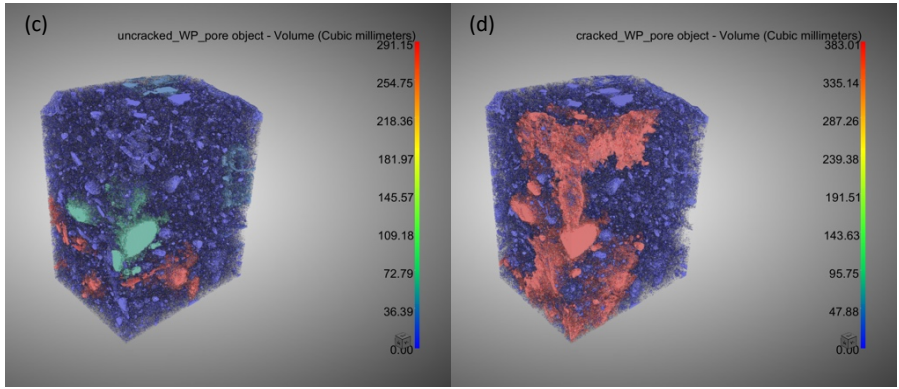

**Figure 12.** 3D rendering of connected objects for the pore phase in the order of increasing volume: (a) uncracked DP sample, (b) cracked DP sample, (c) uncracked WP sample, (d) cracked WP sample.

Fig. 12 presents visualizations of the 3D pore connectivity of both DP and WP and shows similar cracking patterns. Only one major macrocrack is observed at the side corners of the DP sample by comparing (b) and (a). Multiple pores that range from 0 to 30 mm$^3$ (i.e., 0 to 1.9 mm if the pores are assumed to be perfectly spherical) within this area are connected by this macrocrack, which significantly increases the connectivity of the pores (from 9.37% to 21.22% for DP) and the water penetration rate within this area. However, the majority of large pore objects outside this area preserve their volume, and the connectivity is weak because the pore connectivity without the largest pore for the cracked DP sample is very low (only 13.3% of the total porosity). This result means that the water penetration rate outside the fracture zone would not vary significantly when the cracks are introduced. Similar results can be observed for the WP sample when comparing (d) and (c). The difference between DP and WP is that the major macrocrack of WP has a larger volume and connects more pores than that in DP. Therefore, the water penetration in cracked WP should be faster than in cracked DP. Figs. 23 and 25 attest to the deductions related to the water penetration experiments.

The discrete pore size histograms in DP and WP for both uncracked and cracked states are statistically shown in Fig. 13, where the number of pore objects in the samples in different size ranges is plotted against the volume diameter. For both DP and WP, the cumulative number of pores with a diameter range of 100 μm - 3 mm decreased when the cracks were introduced. In the diameter range 4 – 10 mm, the cracked samples contain more pores than the uncracked samples. The largest pores (diameter > 4 mm) that were originally in the uncracked samples disappeared as individual objects and were replaced by much larger pores (diameter > 6 mm). The visualization results in Fig. 12 and Table 2 suggest that the major macrocrack is likely to originate mainly from the largest pores and that these cracks connect to pores that were originally disconnected.

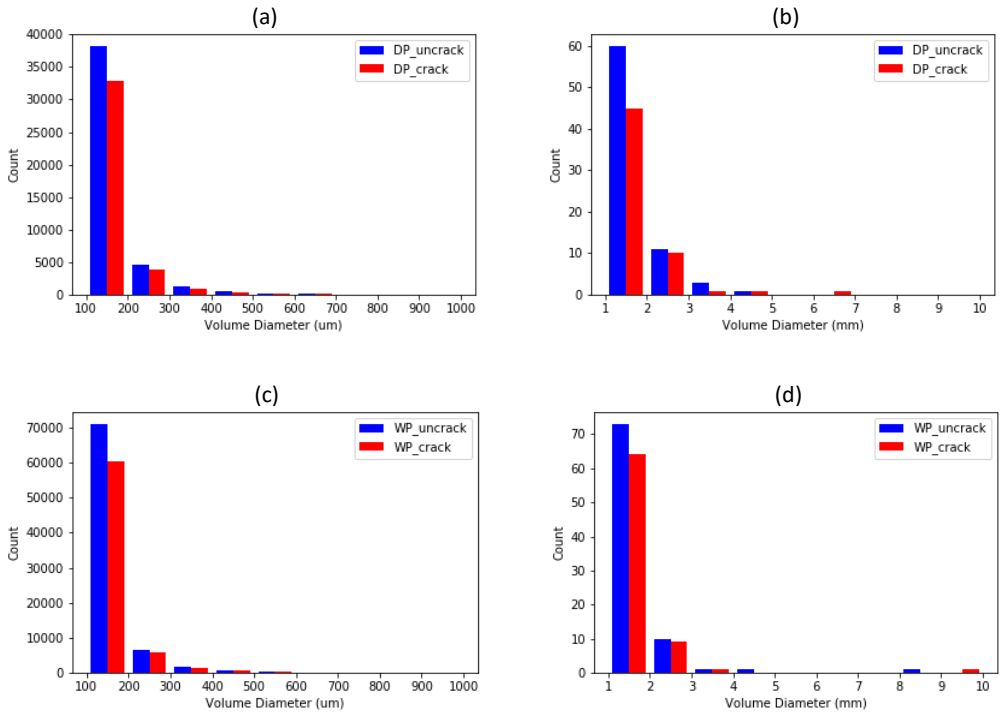

**Figure 13.** Discrete pore size histograms of DP sample (a, b) and WP sample (c, d). For (a) and (c) plots, the diameter range is from 100 μm to 1000 μm. For (b) and (d) plots, the diameter range is from 1 mm to 1000 mm. The left plots characterize the small pores, while the right plots characterize the large pores.

The 3D PSDs were computed by the DPSD (Fig. 14 a and b) and CPSD (Fig. 14 c and d) methods. The 3D PSD is sensitive to the measurement method [44]. For both DP and WP, 3D DPSD provides a broad distribution of pore diameters from 36 μm to 8712 μm. The distribution increases significantly for diameters larger than 3000 μm, and these large pore objects describe a large cumulative pore volume fraction. The comparison between the dashed curves (the cracked sample) and solid curves (the uncracked sample) shows that the pore volume fractions for diameters larger than 3000 μm increased significantly after introducing cracks. In comparison, the pore volume fractions for diameters smaller than 300 μm decreased, which is similar to the results from Figs. 12 and 13. However, with the DPSD method, any pore volume is represented by a single sphere even if the pore volume is composed of multiple connected pores and throats. Therefore, the DPSD is not a reliable method for providing the PSD and usually exaggerates the diameter of the pore volume, especially the large connected pore volume, yet it provides insight on pore volume repartition as a first approach.

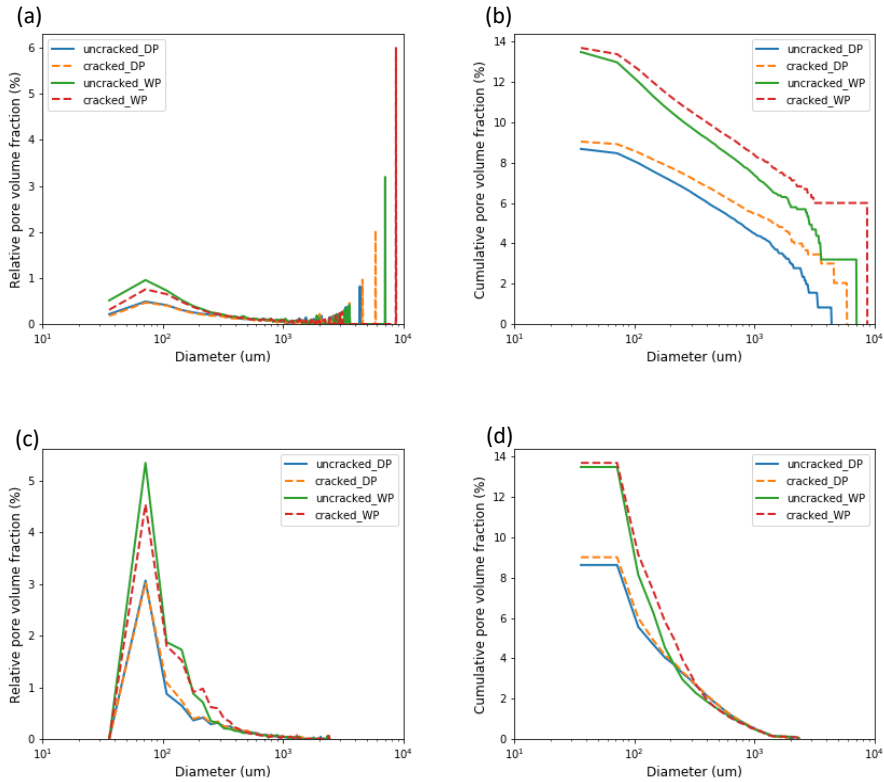

**Figure 14.** Combination of pore size distribution (PSD) of 3D segmented pore volumes, as obtained by (a) 3D relative discrete PSD, (b) 3D cumulative discrete PSD, (c) 3D relative continuous PSD, (b) 3D cumulative continuous PSD.

The 3D CPSD accounts for the ability of a fluid to access a complex pore shape of a given size and avoids the "ink bottle effect", as with MIP [63, 64]. For both Pompeii samples with complex elongated pores, 3D CPSD gives a more reliable PSD and peak pore size. From Fig. 14 c and d, the 3D CPSD provides a small distribution of pore diameters from 36 μm to 2400 μm, with a peak diameter at 50-100 μm. The distribution decreases for diameters that are larger than 100 μm. Comparing the CPSDs of cracked samples and uncracked samples, the pore volume fractions of pore diameters from 100 μm to 170 μm increased slightly after introducing cracks into the DP. For WP, the increments in the pore volume fractions after cracking occurred at pore diameters from 170 μm to 400 μm. Note that the diameter corresponding to the increment essentially indicates the width of the introduced crack in the cracked DP and WP since the pore size in CPSD is quantified by the Euclidean distance to the nearest boundary. Therefore, the widths of introduced cracks in the ancient Pompeii samples ranged from approximately 100 μm to 400 μm.

When comparing the total number of pores, the porosity, and the pore size distribution in the uncracked WP and uncracked DP, it is interesting to note that the uncracked WP has a larger porosity and a greater number of pores within any range from 100 μm to 3 mm. Moreover, the introduced cracks in the WP are wider than those in the DP. The difference can be attributed to the variability from one sample to the other and/or to small cracks induced by capillary effects during the air-drying of the DP sample.

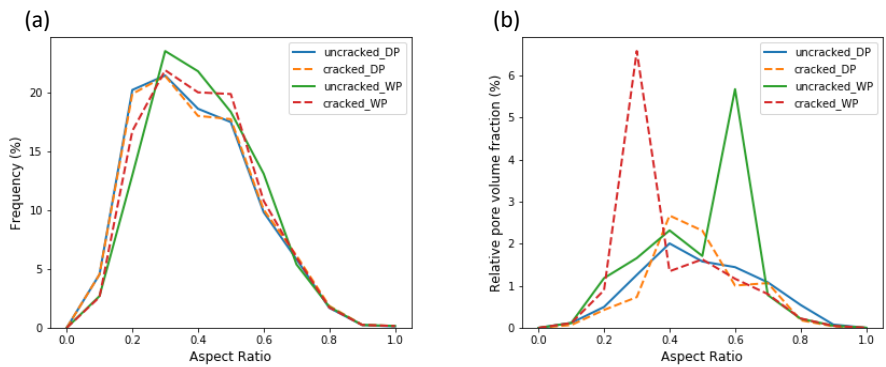

**Figure 15.** Distributions of the 3D aspect ratio (elongation): (a) aspect ratio versus frequency; (b) aspect ratio versus relative pore volume fraction.

Fig. 15 shows the aspect ratio distributions, which statistically describe the shapes of pores in different samples when they are modeled as ellipses. From Fig. 15 a, the peak aspect ratio for both DP and WP samples ranges from 0.2 to 0.4, which corresponds to rather elongated pores (the smallest ellipse eigenvalue is 0.2 to 0.4 smaller than the largest eigenvalue). The difference between the curves of the uncracked and cracked samples is small, which suggests that the introduced cracks affect only the shape of a very small number of pores. However, Fig. 15 b shows that this small number of pores accounts for the majority of the volume fraction. For sample WP, the peak aspect ratio that corresponds to the pore objects of ~6% volume fraction (~46% of total porosity) decreased from 0.6 to 0.3 after introducing the cracks. Similar results are obtained for DP. After introducing the cracks, the volume fraction of the elongated pores with a peak aspect ratio of 0.4 increased from ~2% (23% of the total porosity) to ~3% (35% of the total porosity). The volume fractions of pores that have aspect ratios close to 1 decreased after introducing cracks into both DP and WP. The above results indicate that the pores become connected by the introduced cracks and merge into more elongated pore networks, considering the small variance in the porosity.

3.3 Neutron radiography results

3.3.1 Water distribution in the originally wet Pompeii sample

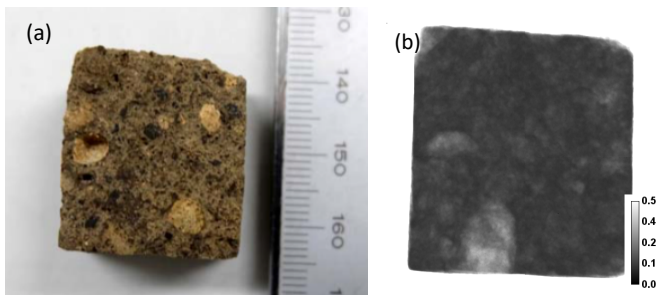

**Figure 16.** Photograph of uncracked sample WP (a) and white spectrum neutron transmission image (b) of uncracked WP measured in its original wet conditions (the sample was not dried intentionally after extraction from the location site). The contrast in (b) is due to attenuation by both concrete and water.

Sample WP was maintained wet since its extraction from the structure. The white spectrum neutron transmission image in Fig. 16 b shows the water distribution in the uncracked WP. Water is evenly distributed throughout the WP, except for minor bright regions. The neutron image (Fig. 16 b) is similar to the 2D porosity density image (Fig. 17 a). The bright regions in Fig. 16 b with high neutron transmission values correspond to the highlighted zones in Fig. 17 a, which have high porosity density (>0.5). The results indicate that the original wet Pompeii sample might have dried out sometime between the time of the sample extraction and the neutron radiography experiment, since it was not sealed in a leak-free container. Indeed, water in the large pores exposed to the air is bound to evaporate first.

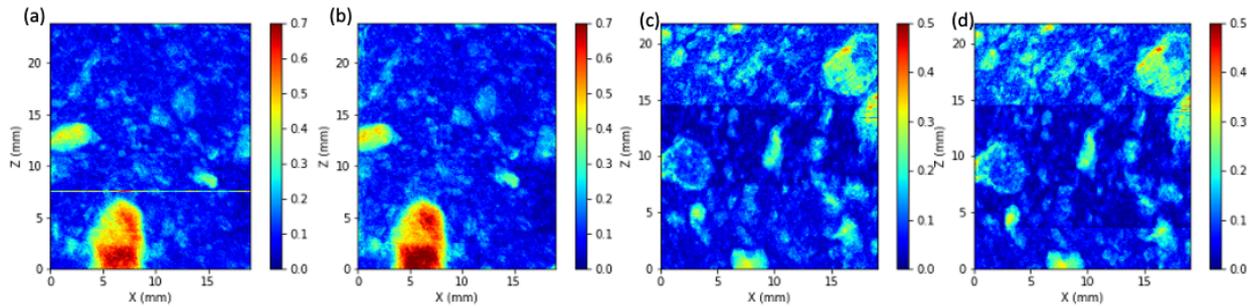

**Figure 17.** 2D porosity projection (XZ plane) of the porosity density in (a) uncracked WP sample, (b) cracked WP sample, (c) uncracked DP sample, and (d) cracked DP sample. The porosity density value at each pixel is defined as the number of pore voxels divided by the total number of voxels along the Y-axis during the projection.

3.3.2 Quantitative map of the water distribution

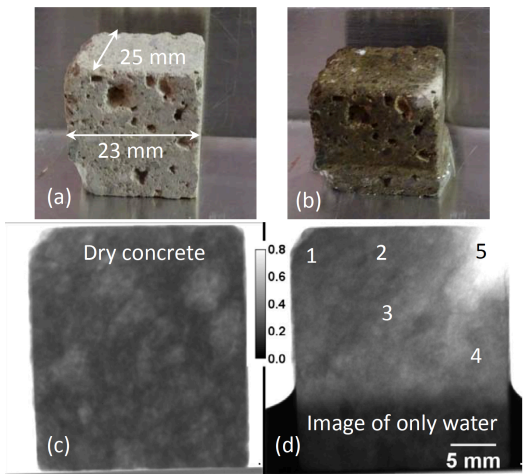

**Figure 18.** Example of step-by-step neutron radiography data analysis: (a) Photograph of the cracked dry concrete sample from Pompeii used in the experiment, (b) The same sample after absorption of water by capillary action (the sample was partially immersed in water at the bottom), (c) White beam neutron transmission of the dry sample, (d) Neutron transmission image of water only (the wet image was normalized by the image of the dry sample). The grayscale bar in (c), (d) indicates the measured transmission values. The numbers in image (d) indicate the areas of 2x2 mm$^2$ for which the measured transmission is shown in Fig. 18.

An example of step-by-step neutron transmission spectral analysis and quantitative mapping of the water distribution is shown in Fig. 18 in the case of cracked DP. The same sample was imaged using μCT. After imaging the dry sample for 17 minutes in the beam, water was added to the base aluminum container. Water was absorbed by the sample due to capillary forces over an ~20-minute period, during which white-spectrum transmission images were acquired to study the dynamics of the water penetration (next section). Then, a set of energy-resolved images of the sample fully saturated by water were subsequently acquired with ~16 minutes of integration. Figure 18 (d) shows the neutron transmission image of the water within the concrete sample, integrated over all energies. In this image, the contribution of the concrete is removed by normalization of the wet image by the image obtained with the same sample in a dry state, taken at the beginning of the experiment (Fig. 18.c).

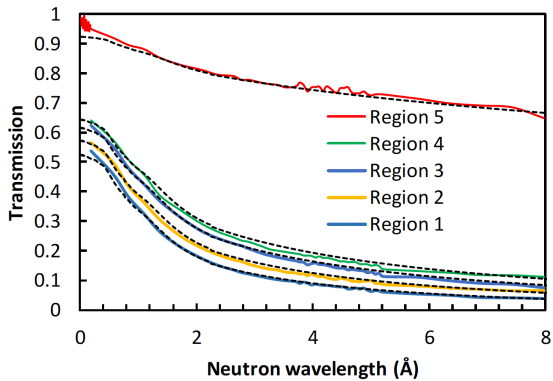

**Figure 19.** Measured (solid lines) and fitted (dashed lines) neutron transmission spectra within the five regions of wet Pompeii concrete sample indicated by the numbers in Fig. 18.

Transmission spectra of the five regions marked in Fig. 18.d are shown in Fig. 19, together with the fitted theoretical transmission based on tabulated cross-sections [65]. The result of this analysis is a water distribution map within the concrete sample, shown in the image of Fig. 20.a, where different colors depict the amount of water integrated through the entire sample thickness. The bottom part of the sample was immersed in water and had no meaning in this reconstruction (white area of the image). The upper right corner did not fully saturate over the 20-minute experiment duration and had the smallest amount of water. The nonuniformity of the water absorption in this particular sample was due to the presence of a diagonal crack, which was intentionally produced by the applied compressive stress load before the neutron imaging experiment.

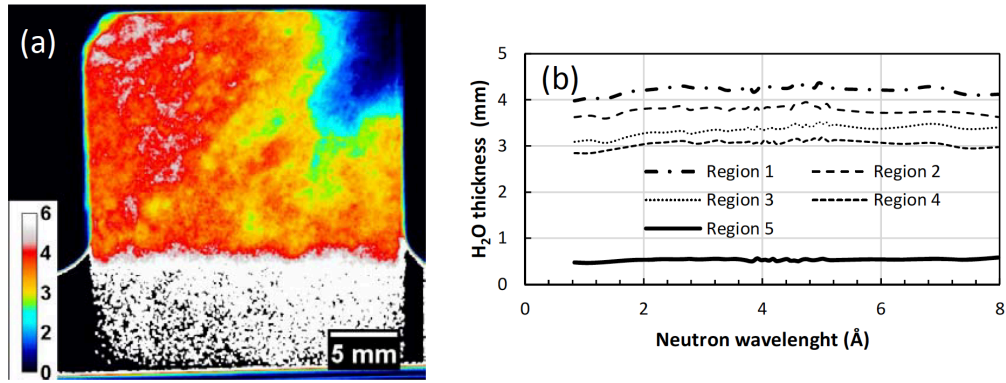

**Figure 20.** (a) Map of the effective water thickness in the dry Pompeii concrete sample reconstructed from the measured neutron transmission spectra. The color bar indicates the water thickness in mm. No data was reconstructed for the bottom part of the sample, which was immersed in water. The right top corner was not saturated with water over the 20-minute test. (b) Water thickness reconstructed from transmission measured at different wavelengths for the five areas of 2x2 mm$^2$ shown by the numbers in Fig. 18.

The range of neutron energies that can be used for water quantification is checked by analyzing narrow-wavelength spectra individually (instead of fitting the entire transmission spectrum). Such an analysis, in most cases, cannot be conducted for each individual pixel because neutron statistics per pixel per narrow wavelength range are rather limited. However, to check the linearity of water quantification, the pixels are grouped within a given region, as was done over the ~2x2 mm$^2$ areas for the results shown in Fig. 20.b. The reconstructed water thickness has a reasonably linear dependence on the neutron wavelength, although some reduction in the water thickness is observed for both short (1-2 Å) and long (7-8 Å) wavelength ranges. These two ends of the neutron spectrum are the most sensitive to the accuracy of background calibration because the neutron flux at these wavelengths was the smallest at the MLF facility in J-PARC. A more accurate background and scatter calibration can improve the linearity of these curves and thus improve the accuracy of the reconstructed water thickness.

3.3.3 Dynamics of water absorption in dry Pompeii sample before and after introducing cracks

Neutron images were taken at constant interval times (typically multiple times per second) from the beginning of the water absorption experiments; thus, the process of water absorption of the DP sample before (Fig. 21) and after (Fig. 22) introducing cracks can be visualized. This process provided the flexibility of using different time scales during the data analysis. The neutron images clearly present both the rise in the water level (Z-axis) and the water penetration along the horizontal direction (X-axis). For the uncracked DP, the water level rose relatively evenly on the horizontal line, as shown in Fig. 21, although the pores are not evenly distributed along the horizontal direction, as shown in Fig. 17 (a).

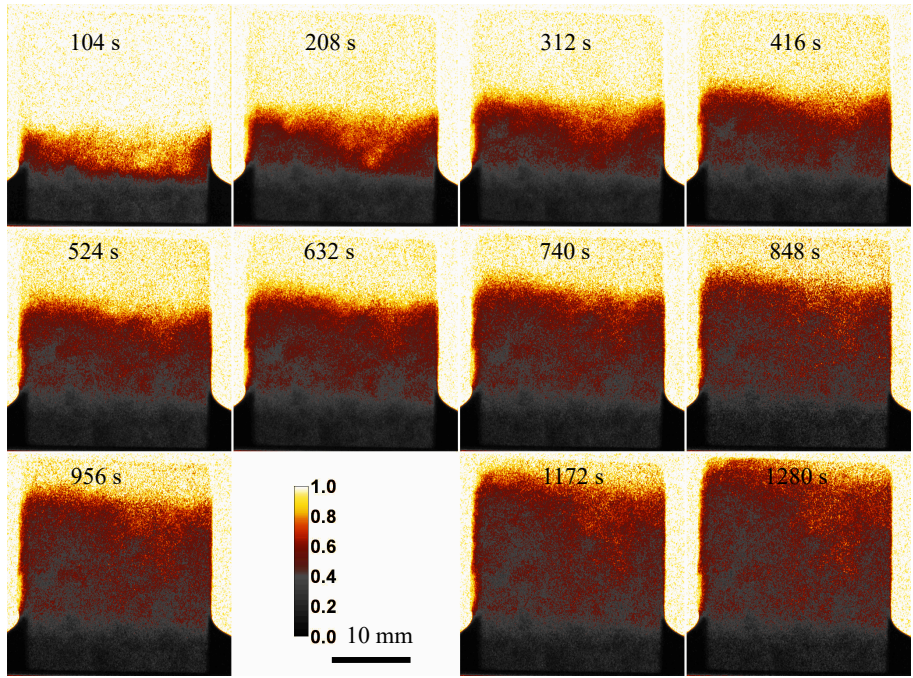

**Figure 21.** Neutron transmission images of uncracked DP during water absorption due to capillary forces. Only attenuation by water is shown (the contribution of concrete is removed by normalization by the image taken with the dry sample). Consecutive images taken with an interval time of 208 s are shown (integration time per image 104 sec). The image at 1064 s is absent due to the interruption in the neutron beam production. The color bar indicates the measured neutron transmission ratio.

Fig. 22 shows the uptake of water in the cracked DP. Because multiple macrocracks and microcracks have been introduced by compression, the uptake of water in the cracked DP is faster than in the uncracked DP. Instead of rising evenly on the horizontal line, the water level in the cracked DP rose much faster in the regions where cracks propagated and formed fracture planes (Fig. 9 f and Fig. 10 f) than in the regions without cracks. Considering the 3D rendering of connected porosity in the cracked DP sample, this result indicates that the rate of water uptake is strongly correlated with the connected pore/crack networks rather than the isolated pores.

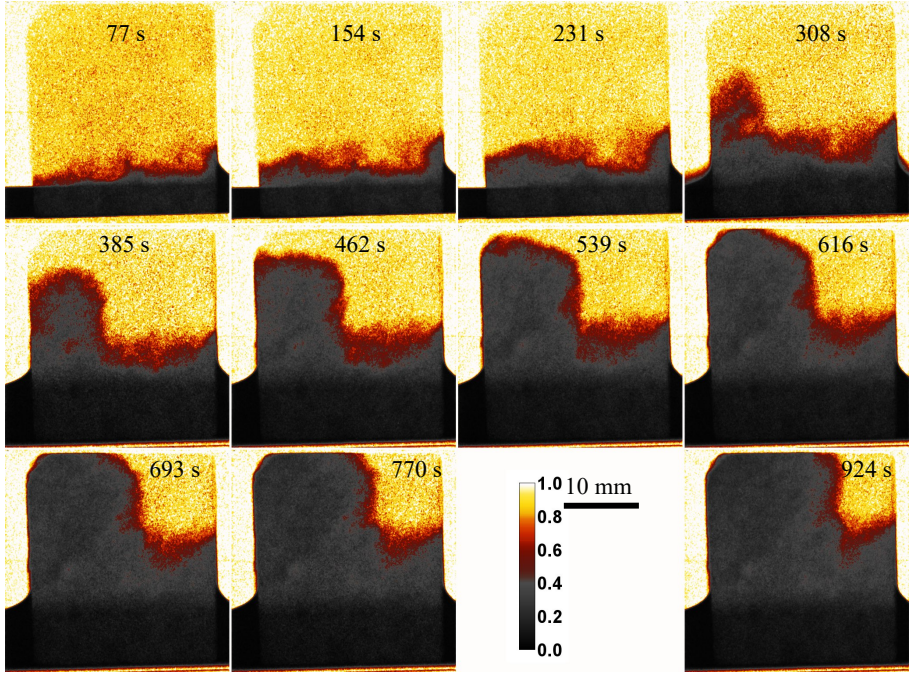

**Figure 22.** Neutron images and water distributions of cracked DP with an interval time of 77 s. The integration time per image is 38.5 sec. The color bar indicates the measured neutron transmission of water only.

From the perspective of a qualitative understanding of the images, previous discussions indicate that the connected pore network plays an important role in the rate of water uptake. Here, Fig. 23 quantitatively evaluates the relationship between the water uptake rate and the cross-sectional area of pores along with the sample height, as follows. Water levels during the water absorption experiments were manually measured using Fiji. For Fig. 23 (d), the total areas of the pore phase along the height are calculated based on the μCT image segmentation results. The water level (absorption depth) h, as a function of time t, can be predicted for ideal capillary absorption by means of Eq. 1. (Note the limitation for h: at a certain height, the water column gravity will equalize the capillary force, and the water level will stop rising).

$$h(t) = Bt^{1/2} \tag{1}$$

where B is the coefficient of capillary penetration (sorptivity). In Fig. 23 b and c, linear regression and segmented regression (four segments) fit the dataset ($t^{1/2}$, h) well, with a coefficient of determination of over 95%. Therefore, the coefficient of capillary penetration B can be approximated as the slope of each curve segment ($t^{1/2}$, h).

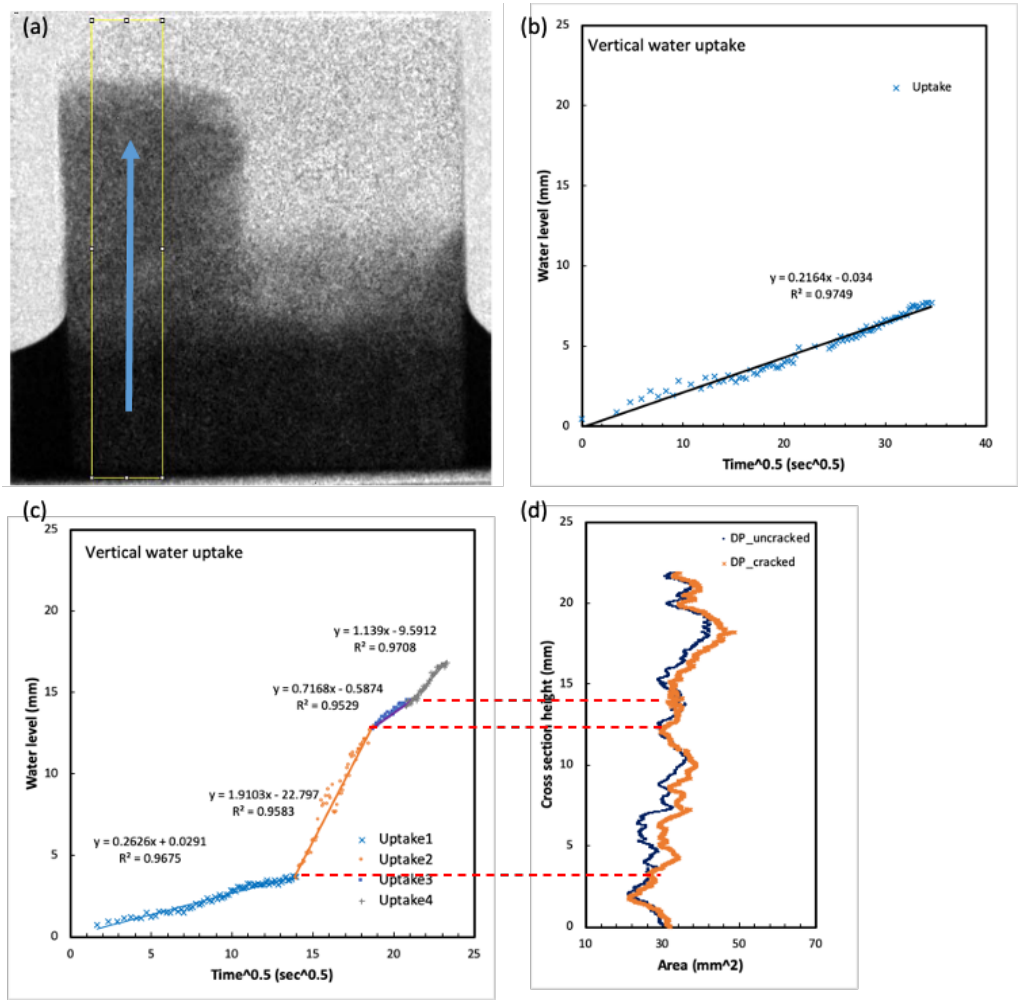

**Figure 23.** Correlation between the rate of water uptake and the cross-sectional area of pores along the height. The water level versus the sqrt of water absorption time is shown in (b) and (c). (a) Shows the locations of the areas along which the dynamics of water absorption are shown in graphs (b) and (c), with (b) uncracked DP, and (c) cracked DP. (d) Total cross section area of the pores given by μCT along the height of sample DP.

For the uncracked sample, the slope of the curve was almost constant during the whole experiment, even though the area of the pores varied along the height. However, the introduced cracks critically influenced the gradients of the segmented curves. As shown in Fig. 23.c, the gradient coefficient B of Uptake1 curve (h < 5 mm) was similar to the gradient of the uncracked sample. When the water level rose to 3 mm, B dramatically increased from 0.26 mm/$\sqrt{s}$ to 1.91 mm/$\sqrt{s}$ due to reaching the region of cracks. The areas of newly introduced cracks were less than 25% of the areas of original pores. This quantitative analysis again indicates that the connected pore/crack networks rather than the isolated pores play an important role in determining the rate of water uptake.

3.3.4 Dynamics of water absorption in dry Pompeii sample after introducing cracks

The dynamic water absorption experiment on the uncracked WP was not conducted since the sample was originally wet. After a first neutron radiography test (to obtain a reference state), the uncracked WP was scanned using μCT and dried out. Then, cracks were introduced into the WP sample. The dynamic water absorption experiment was conducted on an initially dry cracked WP.

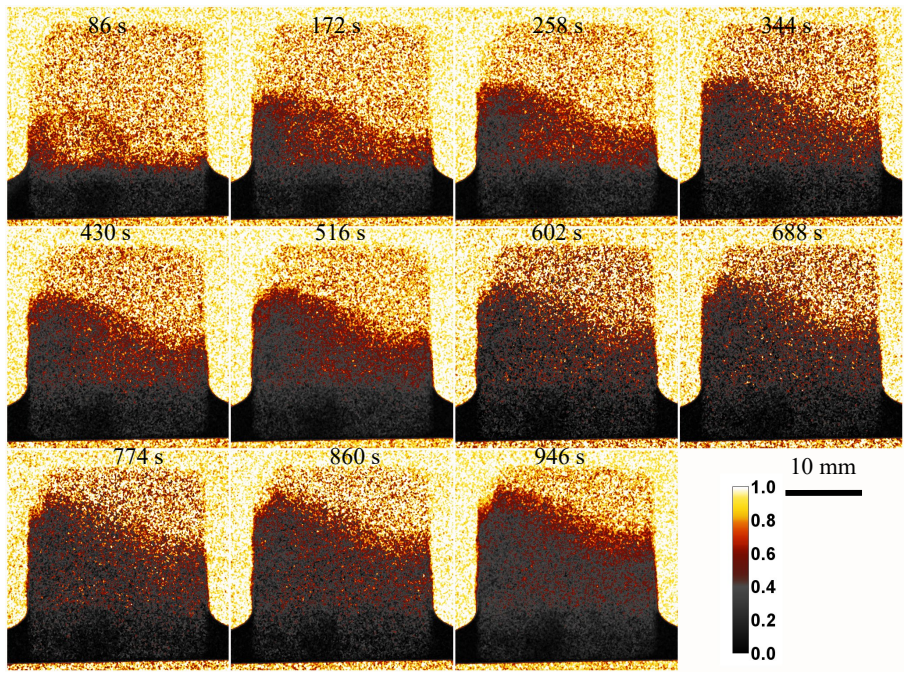

**Figure 24.** Neutron images of the water distribution within the cracked WP taken with an interval time of 86 s. The color bar indicates the measured neutron transmission ratio.

Fig. 24 shows the uptake of water in the cracked WP sample. The uptake of water on the left-hand side was faster than that on the right-hand side. The water level in the cracked DP sample rose much faster in the region $\{X, Z: 0 \text{ mm} \leq X \leq 10 \text{ mm}, 0 \text{ mm} \leq Z \leq 25 \text{ mm}\}$ (where cracks propagated and formed fracture planes (Fig. 11 d and Fig. 11 f)) than in the regions without cracks. Considering the 3D rendering of the connected porosity in the cracked WP sample (Fig. 12 d), this result confirms that the rate of water uptake is more strongly correlated with the connected pore/crack networks than with the isolated pores.

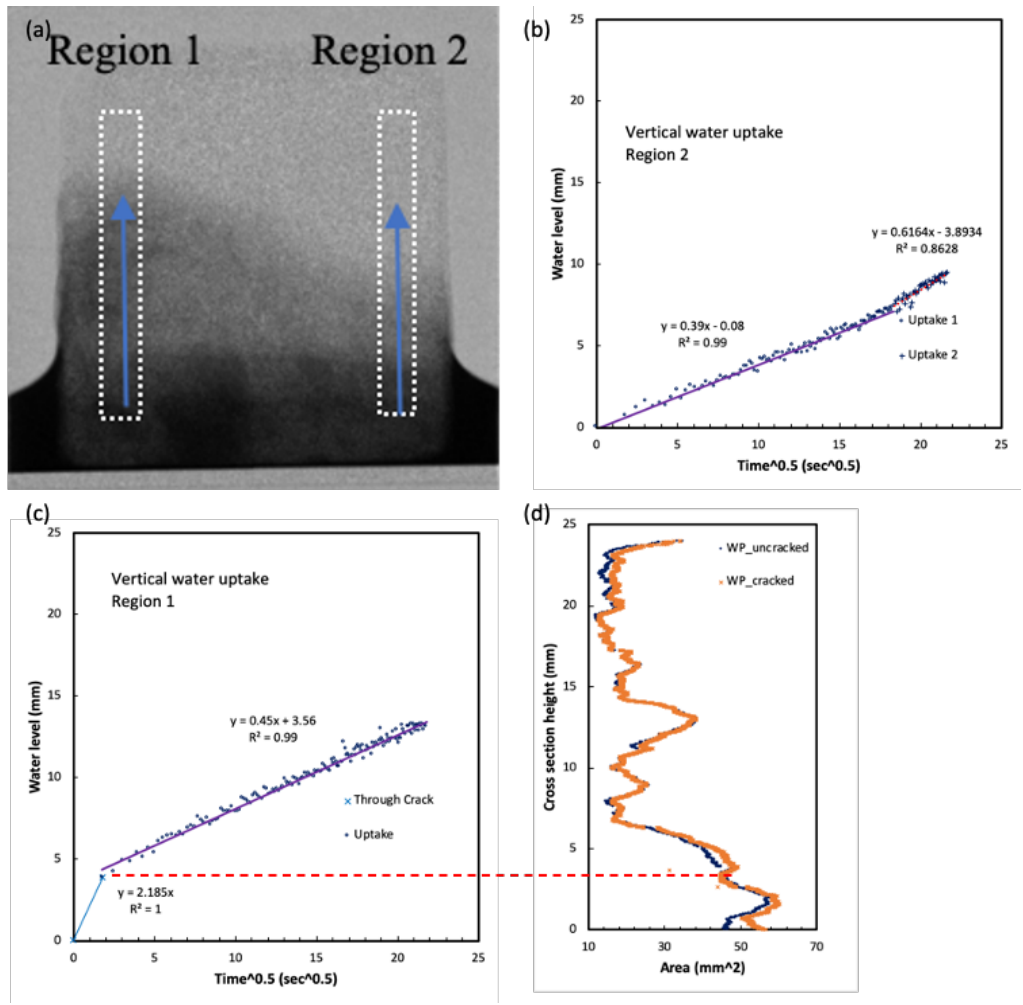

**Figure 25.** Correlation between the rate of water uptake and the cross-sectional area of pores along the height of the cracked WP sample. The water level versus the sqrt of water absorption time is shown in (b) and (c). (a) Location of areas along which the dynamics of water absorption are shown in graphs (b) and (c) for regions 2 and 1, respectively. Note that the water levels in (b) and (c) start at ~10 mm above the sample bottom. (d) Total cross-sectional area of pores along the height of the sample WP.

Similar to DP, Fig. 25 provides a quantitative evaluation of the relations between the water uptake rate and the cross-sectional area of pores along the sample height for the cracked WP sample. For the water uptake on the left-hand side (region 1), the coefficient of capillary penetration (B) is 2.19 mm/√s at first (h < 5 mm) due to the introduced macrocracks and the connected large pore network at the bottom of Region 1. However, B is 0.39 mm/√s at the bottom of Region 2 for the same WP sample. This finding is close to the value of the uncracked DP, which means that this part of the water flux is controlled by similar pore systems. Because of the 3D rendering of the connected objects for the pore phase in Fig. 12 and the significant difference in the pore area below 5 mm in Fig. 25 d, the results indicate that both the connectivity and the volume at the bottom of Region 1 increased after cracking, while the pores in Region 2 preserve their connectivity and size. The macrocracks only initiated and propagated in Region 1 of the WP.

In summary, the local microstructural information, such as the presence of cracks and the increase in the regional pore connectivity, has a significant influence on the water uptake rate. This fact increases the local coefficient of capillary penetration in ancient Pompeii samples. Between the cracked and uncracked regions, the local coefficient of capillary penetration B is different by one order of magnitude. Since this coefficient is proportional to the critical pore size (Eq. 2), through which water penetrates by capillary suction, the order of magnitude difference means that the pore size, which controls the water uptake, is possibly one order of magnitude smaller in uncracked concrete compared to cracked material (Table 6).

3.4 Comparison of the results

3.4.1 Porosity and connectivity

X-ray microtomography [66] has been broadly applied to characterize the 3D structures of paste and concrete, for example, enabling the calculation of porosity and connectivity, which vary as a function of mixture design, voxel resolution, segmentation method, and other minor factors. Using 3D µCT of cement paste with different resolutions, Gallucci et al. [67] reported that an increase in the voxel size (i.e., a decrease in the resolution) (from 0.67 to 1.34, 2 and 2.67 µm) induces a decrease in the calculated porosity (from 18.60 to 11.48, 6.63, and 5.03%, respectively) and in calculated pore connectivity (from 95% to 82, 66 and 0). Table 3 lists the porosity and pore connectivity of typical cementitious materials calculated from microtomography data. Voxel sizes range from 1 µm to 15 µm, and gray level threshold methods were applied for phase segmentation.

The porosity of the paste alone for the ancient Pompeii sample was calculated from segmented µCT images. The porosity of ancient Pompeii paste was defined as the volume of pores surrounded by paste (matrix) divided by the sum of the paste (matrix) volume and the volume of pores surrounded by paste (matrix). On the other hand, the aggregate phase and pore phase within the aggregates were not counted for the calculation of paste porosity. To separate the aggregates and pores within the aggregates from the objects of other phases (matrix and pore connected to matrix), the voxels of pores surrounded by the aggregates merged into the voxels of surrounding aggregates using the hole-filling algorithm in Fiji. Then, the voxels of the merged aggregated phase were deleted from the whole segmented dataset, and only the voxels of the paste (matrix) phase and the voxels of the pore connected to the paste (matrix) phase were left to calculate the porosity in the paste.

**Table 3.** A comparison of porosity for different concretes from microtomography (PC: Portland cement, SHCC: strain hardening cement composites).

| Material | Cement paste volume (% total volume) | Porosity | Pore connectivity (% of total porosity) | Voxel size (um) |
| --- | --- | --- | --- | --- |
| uncracked DP concrete | 79 | 8.68% | 0% (9.37%) | 9.5 |
| uncracked DP paste | 100 | 9.51% | - | 9.5 |
| uncracked WP concrete | 62 | 13.48% | 0% (31.48%) | 9.5 |
| uncracked WP paste | 100 | 16.74% | - | 9.5 |
| PC paste (W/C=0.25) [68] | 100 | 1.36% [68] | - | 5.0 [68] |
| PC paste (W/C=0.5) [67] [47] | 100 [67], 100 [47] | 1.28% [67], 1.8% [47] | -, 14% [47] | 5.0 [67], 1.8 [47] |
| PC concrete (W/C=0.35) [69] | ~ 51 [69] | 2.87% [69] | - | 15.2 [69] |
| PC concrete (W/C=0.5) [70] | ~ 46 [70] | 6.50% [70] | - | 6.18 [70] |
| PC SHCC [71] | 77-87 [71] | 2.4% - 3% [71] | - | 4.8 [71] |

As shown in Table 3, the ancient Pompeii samples have much larger porosities when compared with other typical Portland cement materials even if the voxel size of the ancient Pompeii samples is relatively large. The porosities of ancient Pompeii paste and concrete are a magnitude greater than the porosities of paste and concrete with low W/C. Note that the pore network connectivity of the DP sample is lower than that of PC paste with a w/c ratio of 0.5, although the DP sample shows much higher porosity.

Both MIP and microtomography were used to measure the porosity values in this research, but they are not comparable. MIP measured pores with sizes of 4 nm – 10 microns, while microtomography characterized pores with diameters larger than 19 microns (voxel size of 9.5 microns). From the MIP results, the total porosity of ancient Pompeii samples ranged from 34.5% to 36.2%. In comparison, the porosity of the DP sample from microtomography is 8.86%. This result is consistent with other studies in that microtomography provides only partial quantification of the pore system of concrete [47, 72]. For future study, 3D FIB/SEM is capable of reaching pores down to tens of nm [59], and in the case of modern HPC, pores of tens of nm percolate significantly through the paste.

3.4.2 Coefficient of capillary penetration

**Table 4.** A comparison of the capillary penetration coefficient (B) for different concretes from neutron radiography (unit: mm/min$^{1/2}$).

| Material | Cement paste volume (% total volume) | B of uncracked sample | B of cracked sample |
| --- | --- | --- | --- |
| DP concrete | 79 | 1.67 | 2.03 – 14.80 |
| WP concrete | 62 | 3.02 | 3.48 – 16.92 |
| PC mortar (W/C=0.36) | 35 [73] | 1.57 [73] | - |
| PC mortar (W/C=0.4) | 27 [74] | 0.39-0.61 [74] | - |
| PC mortar (W/C=0.42) | 35 [73] | 2.24 [73] | - |
| PC mortar (W/C=0.5) | 35 [73] | 2.38 [73] | - |
| PC mortar (W/C=0.6) | 30 [74] | 1.16 [74] | - |
| PC concrete (W/C=0.4) | 24 [75] | 0.89 [75] | - |
| PC concrete (W/C=0.6) | 24 [75] | 1.43 [75] | - |
| PC SHCC | 63 [76] | 1.14 [76] | - |

When comparing the capillary penetration coefficient B of ancient Pompeii samples and conventional cement materials in Table 4, it is interesting to see that the capillary penetration coefficient of DP and WP are similar to modern Portland-based reference cement materials, although the porosity of the ancient Pompeii sample is much higher. Therefore, the porous ancient Pompeii mortar/concrete has a comparable capillary penetration rate to modern Portland cement materials, and the capillary penetration rate is key to the resistance to long-term erosion of water and to improved durability. Similar water penetration rates also mean that the percolating pore systems of both ancient Pompeii and modern Portland cement-based materials have similar pore sizes, and these are the pores that control fluid ingress, permeability, and durability.

3.4.3 Permeability prediction

The capillary penetration coefficient B is related to a critical pore size $d_c$ by the Washburn equation:

$$B = \sqrt{\frac{\gamma d_c \cos(\phi)}{4\eta}} \qquad (2)$$

where $\gamma$ is the interface surface tension, $\eta$ is the dynamic viscosity, and $\phi$ is the contact angle. For undamaged ancient Pompeii concrete, critical pore sizes $d_c$, which correspond to the measured penetration coefficients B, are on the order of 2.6-8.2 nm (Table 5).

**Table 5.** Prediction of critical pore sizes using the Washburn equation along the Z-axis (vertical) direction. Assuming that at room temperature (25 °C), the water/cement/air interface surface tension $\gamma$ is 66 mN/m [77], the dynamic viscosity $\eta$ of water is 0.89 mPa.s, and $\cos(\phi)$ is 1 (water is a perfectly wetting fluid) [78].

| Sample | B (mm/$\sqrt{s}$) | $d_c$ (nm) |
|---|---|---|
| uncracked DP sample | 0.22 | 2.6 |
| cracked DP sample | 0.26-1.91 | 3.8-196.8 |
| cracked WP sample | 0.39-2.19 | 8.2-258 |

These values are consistent with the typical pore sizes of modern Portland cement paste hydrates (C-A-S-H) [57]. They also describe the main channels for water flow, i.e., they describe a percolating pore system.

In a first approach, a 1D Katz-Thompson equation [62] can describe the fluid permeability K as a function of a critical pore size $d_c$, which is characteristic of fluid percolation, as follows:

$$K = \frac{d_c^2 \phi_{percolating}}{226\tau} \qquad (3)$$

This formula yields the values in Table 6 for K, which are on the order of $10^{-20}$-$10^{-19}$ m² for ancient Pompeii concrete with the $d_c$ from Table 5. Such values correspond to those of highly durable modern Portland concretes [59, 79], which explains the actual durability of ancient Pompeii concrete. In addition, the predicted K varied with $d_c$ from different measurements. Further experiments for the permeability measurement are required.

For cracked Pompeii concrete, however, Poiseuille's law (Eq. 4, where $h_{min}$ is the minimum width of the percolating pore/crack) [80] is more adaptable (see Table 7). This approach yields permeability values that are greater by several orders of magnitude than those of undamaged concrete.

$$K = \frac{h_{min}^2}{12} \qquad (4)$$

**Table 6.** Prediction of the fluid permeability using the Katz-Thompson equation along the Z-axis (vertical) direction for the uncracked Pompeii sample.

| Sample | Porosity | $d_c$ (nm) | Tortuosity | $K$ (m$^2$) |
|---|---|---|---|---|
| uncracked Pompeii mortar fragments | 34.5%-36.2% (MIP) | 770-860 (MIP) 20-40 (modern PC) | 1.25-1.32 (μCT) | 6.9-9.5 e$^{-16}$ 4.6 e$^{-18}$-2.1 e$^{-17}$ |
| uncracked DP sample | | 2.6 (neutron) | | 1.0-1.2 e$^{-20}$ |
| cracked DP sample | | 3.8-196.8 (neutron) | | 1.9 e$^{-20}$-4.9 e$^{-17}$ |
| cracked WP sample | | 8.2-258 (neutron) | | 7.8 e$^{-20}$-8.5 e$^{-17}$ |

**Table 7.** Prediction of fluid permeability using Poiseuille's law, along the Z-axis (vertical) direction, for the cracked Pompeii sample.

| Sample | $h_{min}$ (μm) | $K$ (m$^2$) |
|---|---|---|
| cracked Pompeii concrete | 19-100 (μCT) | 3.0 e$^{-11}$-8.3 e$^{-10}$ |

## 4. Conclusions

Integrated synchrotron X-ray μCT and neutron radiography with computer vision analysis is a powerful nondestructive testing method for conducting a systematic investigation of the complex concrete microstructure and the dynamics of water penetration. Synchrotron X-ray microtomography provides high-resolution 3D microtomography images with a broad field of view. After the reconstruction of the μCT image, machine learning-based image segmentation methods and DVC analysis acquired both qualitative and quantitative microstructural information from the ancient Pompeii concrete samples. An innovative pipeline was developed to process the associated large datasets autonomously and accurately. Segmented pore and aggregate phases were visualized in 3D rendering. Multiple microcrack and macrocrack propagations, and a wider dispersion of cracks in both dry and wet Pompeii samples were observed, which presented a stable and ductile fracture pattern. 3D morphological and statistical microstructure analysis from the accurate segmented microtomography image provided 3D pore network information, such as porosity, pore size distribution, and pore connectivity, which explain the differences in the water uptake rates at different mesostructural regions in dynamic neutron radiography experiments.

The experimental results show that the water penetration rate depends highly on the local microstructure, such as the regional pore connectivity, and the presence of cracks, while it depends less on global microstructural information, such as global porosity. For uncracked ancient Pompeii samples, the water capillary penetration coefficients (1.67 and 3.02 mm/min$^{1/2}$) in the dry and wet samples, respectively, are both comparable to modern PC paste/concrete (which range from 0.39 to 2.38 mm/min$^{1/2}$). At the same time, ancient Pompeii concrete (8.86% and 18.22% for dry and wet samples, respectively, based on microtomography and a range of 34.5% to 36.2% based on MIP) is more porous than modern PC paste/concrete (W/C = 0.5, with a range from 1.28% to 6.50% based on microtomography and a range from 27% to 31% based on MIP [72, 81]). Note that the porosity calculated from microtomography here ignored the smallest pores, which represent ~ 75% of the whole pore volume. After introducing cracks, the global porosity for both samples is almost the same, but the pore connectivity increases. The pores become connected by the introduced cracks and merge into more elongated pore networks. Instead of being a constant, the water uptake rate and capillary penetration coefficient varied depending on the local microstructure. The capillary penetration coefficient in regions with crack propagations increased significantly (>10 mm/min$^{1/2}$), while the capillary penetration coefficient in the uncracked regions remained similar. Although further experiments for permeability measurements are required, the predicted permeability of uncracked Pompeii samples based on the 1D Katz-Thompson model and Washburn equation ranges from $10^{-20}$ to $10^{-19}$ m$^2$, which corresponds to that of highly durable modern Portland concretes.

The high durability of porous ancient Pompeii concrete benefits from the ductile fracture pattern and low connectivity in the pore network. Integrated synchrotron X-ray microtomography and neutron radiography with computer vision analysis is a step forward in the understanding of the durability of ancient Pompeii concrete and other complex cement materials.

**Acknowledgments**

This work was supported by the SusChEM program, Grant No. DMR-1410557, and the Division of Materials Research Ceramics Program, DMR-CER, Grant No.1935604 of National Science Foundation. This research used resources (Beamline 8.3.2) of the Advanced Light Source, a U.S. DOE Office of Science User Facility under contract no. DE-AC02-05CH11231. The authors thank Dilworth Parkinson for assistance with the μCT experiments. We also acknowledge access to the neutron imaging facility at J-PARC and the great help of our colleagues Dr. T. Shinohara and K. Oikawa with the neutron imaging experiments. These neutron experiments at the Materials and Life Science Experimental Facility of J-PARC were performed under a user program (Proposal No. 2016B0183).